%% file: main.tex
\documentclass[acmsmall, screen,nonacm]{acmart}

\usepackage{adjustbox}
\usepackage[normalem]{ulem}
\usepackage{tabularray}
\usepackage{multirow}
\usepackage{colortbl}
\usepackage{hhline}
\usepackage{rotating}
\usepackage{mathtools}
\usepackage{xspace}
\usepackage[noabbrev,capitalise]{cleveref}
\usepackage{subcaption}
\usepackage{graphicx}
\usepackage{enumitem}

\usepackage{amsmath}
\usepackage{tikz}
\usepackage{pgfplots}
\usepackage{hyperref}

\usepackage{graphicx}
\usepackage{float}
\usepackage{caption}
\usetikzlibrary{positioning,fit,shapes.geometric,backgrounds,calc}
\pgfplotsset{compat=1.18}

\usepackage{todonotes}
\presetkeys%
    {todonotes}%
    {inline,backgroundcolor=yellow}{}

\def\addlegendimage{\csname pgfplots@addlegendimage\endcsname}
\definecolor{forestgreen(web)}{rgb}{0.13, 0.55, 0.13}
\definecolor{lightgray}{rgb}{0.83, 0.83, 0.83}

\AtBeginDocument{%
  }

\acmConference[]{Pre-print}{under review}{September 2024}
\renewcommand\footnotetextcopyrightpermission[1]{} %
\setcopyright{none}

\begin{document}

\title{Fashion Image-to-Image Translation for Complementary Item Retrieval}

\author{Matteo Attimonelli}
\orcid{0009-0003-6600-1938}
\email{matteo.attimonelli@poliba.it}
\affiliation{%
  \institution{Politecnico Di Bari}
  \city{Bari}
  \country{Italy}
}

\author{Claudio Pomo}
\orcid{0000-0001-5206-3909}
\affiliation{%
  \institution{Politecnico Di Bari}
  \city{Bari}
  \country{Italy}}
\email{claudio.pomo@poliba.it}

\author{Dietmar Jannach}
\orcid{0000-0002-4698-8507}
\affiliation{%
  \institution{University of Klagenfurt}
  \city{Klagenfurt}
  \country{Austria}
}
\email{dietmar.jannach@aau.at}

\author{Tommaso Di Noia}
\orcid{0000-0002-0939-5462}
\affiliation{%
  \institution{Politecnico Di Bari}
  \city{Bari}
  \country{Italy}}
\email{tommaso.dinoia@poliba.it}

\renewcommand{\shortauthors}{Attimonelli et al.}

\def\cgcrmlong{\textsc{Generative Compatibility Model (GeCo)}\xspace}
\def\cgcrm{\textsc{GeCo}\xspace}
\def\cgcrmname{\textsc{Generative Compatibility Model}\xspace}
\def\unet{UNet\xspace}

\def\cigm{\textsc{CIGM}\xspace}
\def\cigmname{\textsc{Complementary
Item Generation Model}\xspace}

\def\pogdataset{Taobao Fashion dataset\xspace}
\newcommand{\ie}{i.e., }
\newcommand{\eg}{e.g., }
\newcommand{\nce}{InfoNCE\xspace}

\begin{abstract}

The increasing demand for online fashion retail has significantly boosted research in fashion compatibility modeling and item retrieval. Researchers focus on facilitating retrieval based on user queries, represented as textual descriptions or reference images, by modeling the compatibility among fashion items. Complementary item retrieval, especially top-bottom retrieval, presents challenges due to the need for precise compatibility modeling and traditional methods, predominantly based on Bayesian Personalized Ranking, often exhibit limited performance. More recently, researchers have focused on integrating generative models into compatibility modeling and complementary item retrieval systems to advance this field. These models produce item images that serve as additional inputs for the compatibility model. Interestingly, in the proposed approaches the quality of the generated images is often overlooked, although it could be crucial for the performance of the models. Moreover, although generative models can be effective, they usually require large amounts of data to achieve optimal performance, which can be problematic in contexts where such amount can be limited.

Aiming to overcome these limitations, we introduce the \textit{Generative Compatibility Model (\cgcrm)}, a two-stage approach that enhances fashion image retrieval through paired image-to-image translation. Initially, the first-stage model, called \textit{Complementary Item Generation Model}, which is based on \textit{Conditional Generative Adversarial Networks}, generates images of the target samples (e.g., bottoms) from seed items (e.g., tops), providing conditioning signals for retrieval. Subsequently, these samples are integrated into \cgcrm, the second stage, refining compatibility modeling and retrieval accuracy. Overall, our evaluations on three datasets demonstrate \cgcrm's superior performance compared to state-of-the-art baselines. Our contributions include: (i) the \cgcrm model leveraging paired image-to-image translation within the Composed Image Retrieval framework, (ii) a comprehensive evaluations on benchmark datasets, and (iii) the release of a new Fashion Taobao dataset tailored for top-bottom retrieval, to foster further research. 
Our code is available at: \url{https://github.com/sisinflab/GeCo}.
\end{abstract}



\received{20 February 2007}
\received[revised]{12 March 2009}
\received[accepted]{5 June 2009}

\maketitle

\input{src/introduction}
\input{src/literature}
\input{src/background}
\input{src/methodology}
\input{src/experiments}
\input{src/conclusions}

\bibliographystyle{ACM-Reference-Format}
\bibliography{references}
\end{document}

%% file: src/introduction.tex
\section{Introduction}
The rapid expansion of e-commerce and the escalating demand for personalized fashion recommendations have catalyzed significant advancements in fashion image retrieval systems~\cite{DBLP:journals/csur/DeldjooNRMPBN24}. These systems are designed to retrieve fashion items based on user queries, which may be represented as textual descriptions or reference images.

In this context, complementary item retrieval~\cite{DBLP:conf/www/BibasSJ23} refers to the task of retrieving items that complement a given seed item, forming a coherent ensemble or set, named outfit. Within the fashion context, this involves identifying items that match well with a specified piece of clothing. For instance, top-bottom retrieval~\cite{DBLP:conf/mm/SongFLLNM17} can be considered a specialization of complementary item retrieval, where the seed item is the top (\eg a t-shirt) and the target item is the bottom (\eg a pant). This task is particularly challenging due to the need to understand and model fashion compatibility, which is inherently subjective and context-dependent.

Retrieving fashion items that go together well necessitates a robust approach capable of evaluating the compatibility of two items. Classical approaches often rely on the Bayesian Personalized Ranking (BPR) framework~\cite{DBLP:conf/mm/SongFLLNM17}, leveraging visual and textual information. BPR allows the optimization of the ranking of items by maximizing the difference in predicted scores between a positive and a negative item pair, tailored specifically for implicit feedback scenarios. Even though the integration of generative models in top-bottom retrieval has proven to be effective~\cite{DBLP:journals/ijon/LiuSCM20,DBLP:conf/ijcai/LiuSRNT020,DBLP:conf/www/LinRCRMR19}, these models are primarily optimized for compatibility modeling and often produce low-quality images of small sizes. Furthermore, generative architecture-based models may require significant amounts of data to ensure that they accurately capture the underlying distribution and compatibility relationships. This makes it difficult to effectively use such models when working with small datasets.

In this paper we present a novel method to enhance fashion compatibility modeling and complementary item retrieval through image-to-image translation, named \textit{Generative Compatibility Model (\cgcrm)}. We design our method as a \textit{two-stage} approach: in the first stage, it employs a Conditional Generative Adversarial Network (cGAN), named \textit{Complementary Item Generation Model (CIGM)}, to generate new target samples (named templates~\cite{DBLP:journals/ijon/LiuSCM20}) based on seed items, which serve as conditioning signals for the retrieval process. The~\cigm model executes an image-to-image translation task, learning a mapping between the seed and target items' distributions, trained on pairs of items known to be compatible.

\begin{figure}[t]
    \centering
    \includegraphics[width=\textwidth]{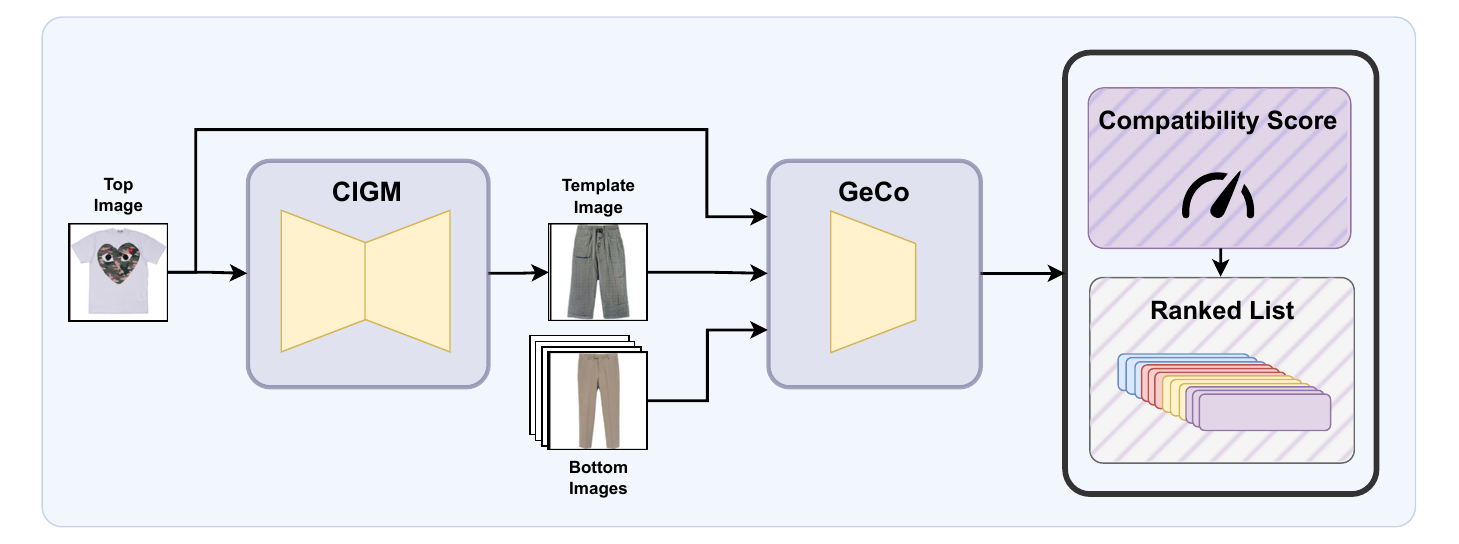}
    \caption{In the proposed architecture the~\cigm model generates bottom templates. Subsequently, the~\cgcrm model leverages the top, the generated template, and the candidate bottom images to evaluate their compatibility. This approach facilitates both compatibility modeling and complementary item retrieval tasks.}
    \label{fig:cgcrm_Workflow}
\end{figure}

In the second stage, \cgcrm incorporates these generated samples to perform compatibility modeling and complementary item retrieval. Specifically, we model these tasks within the Composed Image Retrieval (CoIR) framework, which is designed to retrieve images using queries composed of multiple elements~\cite{DBLP:journals/tomccap/BaldratiBUB24, DBLP:journals/corr/abs-2312-12273, DBLP:conf/iccv/0002OTG21, feng2024improving}. While previous works model the query by combining an input image and a textual description of the desired outcomes, our approach formulates the query by considering both the top and the template images. This results in a more precise modeling of item compatibility and improved retrieval performance. The structure of our model is depicted in \Cref{fig:cgcrm_Workflow}. By placing greater importance on the generated images compared to previous works in the fashion domain~\cite{DBLP:journals/ijon/LiuSCM20, DBLP:conf/ijcai/LiuSRNT020, DBLP:conf/www/LinRCRMR19}, this formulation motivates us to rely solely on visual information for the \cigm and \cgcrm models, distinguishing our approach from prior studies.
Moreover, our proposed two-stage architecture is designed to be efficient and \textit{requires less training data} compared to existing approaches, because it does not require an end-to-end-training.
To the best of our knowledge, while prior works in CoIR have predominantly focused on creating joint queries combining images with textual descriptions~\cite{DBLP:conf/iccv/BaldratiA0B23, DBLP:conf/cvpr/BaldratiBUB22a}, our work is the first to model CoIR using top images and templates as queries, performing compatibility modeling and complementary item retrieval.

Our key contributions are as follows:
\begin{itemize}
    \item We propose the \textbf{\textsc{Generative Compatibility Model} (\cgcrm)}, leveraging paired image-to-image translation within the CoIR framework. Thereby, we enhance the modeling of item compatibility, resulting in more relevant outcomes.
    \item We assess the performance of the model via a \textbf{comprehensive evaluation}: We conduct extensive experiments on benchmark fashion datasets to showcase the effectiveness of our approach, comparing it against several state-of-the-art baselines. 
    \item We release a new version of the Fashion Taobao dataset, named \textbf{FashionTaobaoTB (\textit{Top-Bottom})}: We create a version of the Fashion Taobao dataset tailored for the top-bottom retrieval task. This new dataset comprises over 80,000 fashion items, organized into distinct training, validation, and test sets. To facilitate further research in this domain, we make the dataset publicly available.
\end{itemize}

In the remainder of this paper, we review related work in fashion image retrieval and generative models. Next, we provide the technical background on generative models, focusing on the architecture and training methodology relevant to understanding the proposed approach. After that, we present the experimental setup and results, emphasizing the performance improvements achieved by our method. Finally, we discuss the paper's findings and limitations and sketch future research directions. 

%% file: src/literature.tex
\section{Related Work}\label{sec:2}
In recent years, the domain of fashion recommendation systems has garnered significant attention due to its high potential impact on e-commerce. The ability to accurately recommend fashion items not only enhances the user experience on online retail but also drives sales and customer engagement. This section reviews the existing literature pertinent to our research, divided into two primary areas: fashion item retrieval \& recommendation, and the application of generative models in retrieval tasks.

\subsection{Fashion Item Retrieval \& Recommendation}

The domain of fashion recommendation encompasses a variety of identifiable tasks~\cite{DBLP:journals/csur/DeldjooNRMPBN24}, among which \textit{fashion item recommendation}, \textit{outfit recommendation}, and \textit{complementary item retrieval} emerge as critical areas of research.

Fashion item recommendation~\cite{DBLP:conf/aaai/HeM16,DBLP:conf/sigir/LiuWW17,DBLP:conf/sigir/ChenZ0NLC17} entails suggesting individual clothing items to users based on diverse factors such as preferences, past behavior, demographics, and contextual information. The primary objective is to aid users in discovering clothing items that align with their tastes, needs, and occasions. This task typically employs recommendation algorithms that analyze user-item interactions, item attributes, user profiles, and other pertinent data to generate personalized recommendations.
Outfit recommendation~\cite{DBLP:conf/icip/Zhan021, DBLP:conf/wacv/SarkarBVLBLM23, DBLP:conf/kdd/ChenHXGGSLPZZ19,DBLP:conf/www/CuiLWZW19, DBLP:conf/icmcs/WangCWL21, DBLP:conf/sigir/LiW0CXC20} extends beyond individual fashion items and focuses on recommending complete outfits or ensembles that match the user's style and preferences. Instead of recommending isolated items, outfit recommendation systems strive to curate cohesive combinations of clothing items, accessories, and footwear that complement each other, creating a stylish and harmonious look. Achieving this task necessitates a profound understanding of fashion aesthetics, trends, and the ability to consider the overall coherence and suitability of the recommended outfits for the user's needs and preferences. A special case of outfit recommendation is complementary item retrieval~\cite{DBLP:conf/www/BibasSJ23, DBLP:conf/cikm/HaoZLDFSW20, DBLP:journals/corr/abs-1908-08327}, which aims to identify a target item that complements a specified seed item. In our context, we designate tops as the seed category and bottoms as the target category, thus denoting this task as top-bottom retrieval.

Despite the inherent challenges of outfit recommendation and top-bottom retrieval, which arise from the complexity of capturing relationships among different clothing items, several studies have proposed effective solutions. While outfit recommender systems often utilize the attention mechanisms~\cite{DBLP:conf/icip/Zhan021, DBLP:conf/wacv/SarkarBVLBLM23, DBLP:conf/kdd/ChenHXGGSLPZZ19} or graph neural networks~\cite{DBLP:conf/www/CuiLWZW19, DBLP:conf/icmcs/WangCWL21, DBLP:conf/sigir/LiW0CXC20}, top-bottom retrieval models are typically based on the BPR framework~\cite{DBLP:conf/uai/RendleFGS09}. An early example is BPR-DAE~\cite{DBLP:conf/mm/SongFLLNM17}, which integrates the BPR framework with features of tops and bottoms to retrieve compatible targets. This model combines visual features extracted by a pre-trained Convolutional Neural Network (CNN) with textual features obtained using a TextCNN. The BPR loss is employed to maximize the similarity between the top and the positive bottom while minimizing the similarity between the top and negative bottoms. GP-BPR~\cite{DBLP:conf/mm/SongHLCXN19} extends this work by incorporating a personalization signal. Another relevant work is represented by~\citet{DBLP:journals/tkde/LinRCRMR20}, who propose NOR, an outfit matching schema. This approach leverages features extracted through a pre-trained CNN and utilizes a Gated Recurrent Unit (GRU) to provide explanations for the outcomes of the model.

\subsection{Generative Models for Top-Bottom Retrieval}

The models discussed in the previous subsection are effective in retrieving top-bottom pairs; however, they tend to retrieve similar bottoms for each associated top, thereby limiting diversity and focusing solely on ensuring compatibility.
In response to this challenge, researchers have focused on enhancing the performance of retrieval models by applying Generative Models within the retrieval domain, including \textit{complementary item retrieval} and, specifically, \textit{top-bottom retrieval}.

Notable contributions in the domain of complementary item retrieval include the work by~\citet{DBLP:conf/icdm/KangFWM17}, who proposed DVBPR, a method that incorporates a Siamese Network and a Generative Adversarial Network (GAN) for compatibility modeling and generation. This method is augmented with a preference maximization module that also includes personalization. Other approaches, such as FARM~\cite{DBLP:conf/www/LinRCRMR19} and MGCM~\cite{DBLP:journals/ijon/LiuSCM20}, generate novel bottoms using advanced architectures. FARM integrates a variational transformer to produce compatible bottoms given a top, leveraging learned latent spaces for compatibility modeling. 
In contrast, MGCM introduces a compatibility schema by harnessing a GAN to generate bottom templates, including them in the compatibility assessment.
Specifically, MGCM considers both item-item and item-template relationships: item-item compatibility refers to the compatibility between the top and candidate bottom features, while item-template compatibility denotes the similarity between the candidate and the template. The inclusion of templates in the compatibility modeling enhances both the compatibility modeling and retrieval task performance. Following a similar approach,~\citet{DBLP:conf/ijcai/LiuSRNT020} propose to assess item-item and item-template compatibilities by employing a Cycle-GAN~\cite{DBLP:conf/iccv/ZhuPIE17} generator for the compatibility module.

None of the aforementioned works specifically assess the quality and realism of the generated outputs produced by the generative component, as they primarily focus on the compatibility module. Consequently, these works do not evaluate whether enhancing realism could improve compatibility and retrieval performance. Furthermore, generative models typically form part of the compatibility pipeline, where optimization is achieved through a combination of different loss signals, resulting in challenges with respect to convergence~\cite{DBLP:conf/ijcnn/Thanh-Tung020,DBLP:conf/icml/MeschederGN18,DBLP:conf/iclr/ArjovskyB17}. Additionally, the generated outputs are typically utilized as a regularization component in the compatibility modeling rather than being incorporated into the query composition process~\cite{DBLP:journals/ijon/LiuSCM20, DBLP:conf/ijcai/LiuSRNT020}. Therefore, it may be beneficial to explore methods for assigning greater importance to the templates in the compatibility assessment process.
To this end, a promising way to integrate templates in the compatibility modeling and complementary item retrieval task is to rely on the Composed Image Retrieval framework. Existing works have focused on the creation of joint queries~\cite{DBLP:journals/tomccap/BaldratiBUB24, DBLP:journals/corr/abs-2312-12273, DBLP:conf/iccv/0002OTG21, feng2024improving}, which combine an image and a textual description of the desired outcome. To the best of our knowledge, only the work by~\citet{DBLP:conf/wacv/TianNB23}
proposes a method for the retrieval of desired fashion images specified by an input image source and a textual description. No existing work investigates the impact of generated templates on the retrieval process, thus outlining an unexplored direction. Aiming to assess the influence of realistic templates in the compatibility assessment, we exclusively consider the visual signal as input to our model. This focus on image-based input underscores our emphasis on the importance of the generated templates, allowing us to evaluate their impact on compatibility assessment without the interference of additional modalities.

As discussed in the subsequent sections, our work explores this direction by effectively addressing both compatibility modeling and complementary item retrieval tasks while still generating realistic templates. Our objective is to enhance the quality of the generated images and integrate them into a CoIR framework, giving the templates the right importance instead of employing generations merely as regularization signals.

%% file: src/background.tex
\section{Technical Background: Generative Models}\label{sec:3}

In this section, we provide the essential technical background to understand the design choices behind our approach. We begin by defining generative models, followed by a detailed description of three pivotal ones: Variational Autoencoders, Generative Adversarial Networks, and Stable Diffusion. 
We discuss the characteristics and limitations of these approaches, which led us to the choice of GANs for our method.

\subsection{Variational Autoencoders}

Generative models are a class of machine learning models that are designed to generate new data samples that resemble a given dataset. Unlike discriminative models, which focus on predicting labels for given data, generative models learn to understand and replicate the underlying distribution of the data itself. This capability makes them particularly powerful for tasks where creating new, realistic data samples is beneficial, such as image synthesis and text generation.

\begin{figure}[t]
    \centering
    \begin{subfigure}[b]{\textwidth}
        \includegraphics[width=\textwidth]{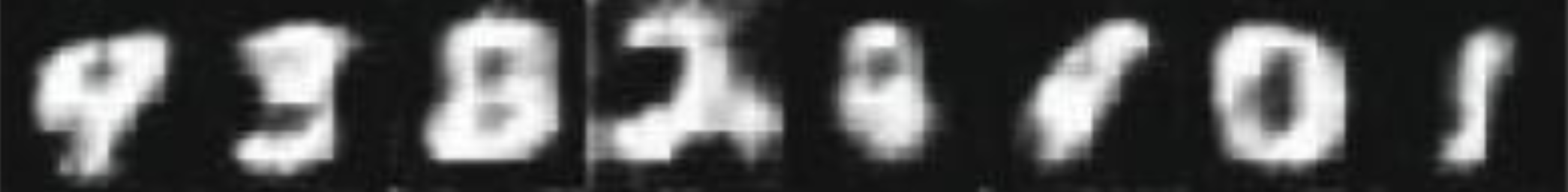}
        \caption{VAE generations.}
        \label{fig:vae_mnist}
    \end{subfigure}
    \hfill
    \begin{subfigure}[b]{\textwidth}
        \includegraphics[width=\textwidth]{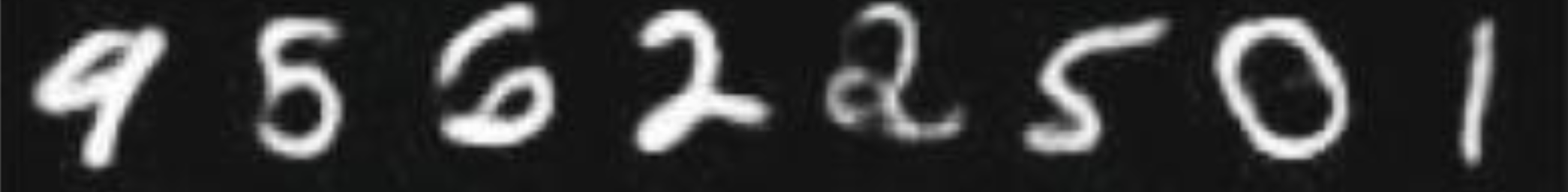}
        \caption{GAN generations.}
        \label{fig:gan_mnist}
    \end{subfigure}
    \caption{An example of generated images from~\cite{DBLP:conf/mspn/El-KaddouryMH19} illustrates the differences in generation quality: (a) presents images generated by a VAE, while (b) showcases images sampled from a GAN.}
    \label{fig:gan_vs_vae}
\end{figure}

The first remarkable work is represented by the introduction of Variational Autoencoders (VAEs)~\cite{DBLP:journals/corr/KingmaW13}. VAEs consist of an encoder, denoted as $q_{\phi}(z|x)$, which maps input data $x$ to a latent space representation $z$. This encoder approximates the posterior distribution $p(z|x)$ using a variational distribution $q_{\phi}(z|x)$, parameterized by $\phi$. The training objective involves maximizing the evidence lower bound (ELBO). The decoder, represented as $p_{\theta}(x|z)$, generates samples $x$ from the latent space $z$, effectively mapping $z$ back to the data space. Although VAEs have demonstrated their effectiveness in various image generation tasks, they often produce blurry samples due to the minimization of the Kullback-Leibler (KL) divergence between the sampled latent representations and the model's prior distribution~\cite{DBLP:journals/corr/Goodfellow17}, as illustrated in \cref{fig:vae_mnist}.

\subsection{Generative Adversarial Networks}
Aiming to enhance the quality of the generated samples, \citet{DBLP:journals/corr/GoodfellowPMXWOCB14} proposed Generative Adversarial Networks (GANs). GANs are a class of generative models comprising two neural networks: the generator and the discriminator, which are trained concurrently through adversarial training. The generator's objective is to create samples that closely resemble real data, while the discriminator's task is to distinguish between real and generated samples. This adversarial training process compels the generator to produce increasingly realistic samples. The generator begins by creating new samples from random noise, typically sampled from a Gaussian distribution. The training of GANs can be formulated as a min-max game with the objective function, shown in \cref{eq:original_gan_obj}, derived from the Jensen-Shannon (JS) divergence, where $p_{data}(\mathbf{x})$ represents the data distribution, and $p_{\mathbf{z}}(\mathbf{z})$ represents the prior distribution over the latent space. The generator $G$ aims to minimize this objective against an adversarial discriminator $D$, which seeks to maximize it. \cref{fig:gan_mnist} showcases some examples of images generated with a GAN on the MNIST dataset, highlighting their realism compared to VAEs ones.

\begin{equation}
    \min_G \max_D V(D, G) = \mathbb{E}_{x \sim p_{data}(x)}[\log D(x)] + \mathbb{E}_{z \sim p_{z}(z)}[\log(1 - D(G(z)))]
    \label{eq:original_gan_obj}
\end{equation}

GANs have been employed in a variety of domains due to their powerful generative capabilities. In computer vision, GANs are widely used for tasks such as creating high-resolution images from low-resolution inputs (super-resolution)~\cite{DBLP:conf/cvpr/LedigTHCCAATTWS17}, generating photorealistic images from textual descriptions~\cite{DBLP:conf/iccv/ZhangXL17}, and generating entire scenes from semantic layouts~\cite{DBLP:conf/cvpr/Park0WZ19}. Initially, \citet{DBLP:journals/corr/GoodfellowPMXWOCB14} framed GANs as models that generate new samples starting from pure noise sampled from a prior distribution, thus enabling the creation of new samples without any control over the actual result. To incorporate a control mechanism over the generated output, researchers have directed their attention to Conditional Generative Adversarial Networks (cGANs)~\cite{DBLP:journals/corr/MirzaO14}. The control signal in cGANs can take various forms, such as textual descriptions~\cite{DBLP:conf/iccv/ZhangXL17} or specified class labels~\cite{DBLP:journals/corr/MirzaO14}, guiding the generation process to produce outputs that adhere to the given conditions. When the control signal is an input image that needs to be transformed from one domain to another, the task is referred to as image-to-image translation~\cite{DBLP:conf/cvpr/IsolaZZE17}. This process involves learning a mapping between two distinct image domains, making cGANs particularly suitable for this application: by conditioning on input images, cGANs can generate corresponding output images. Thus, generating complementary items, specifically top-bottom ones, can be framed as an image-to-image translation task where the generator learns a complex mapping between the two distributions.

In order to effectively address image-to-image translation, the objective function for a cGAN can be formulated as stated in \cref{eq:pix2pix_objective}, where $\mathbf{x}$ denotes the input image and $\mathbf{y}$ denotes the output image.
\begin{equation}
\label{eq:pix2pix_objective}
\min_G \max_D V(D, G) = \mathbb{E}_{x, y \sim p_{data}(x, y)}[\log D(x, y)] + \mathbb{E}_{x \sim p_{data}(x), z \sim p_{z}(z)}[\log(1 - D(x, G(x, z)))],
\end{equation}
The generator $G$ is tasked with generating $\mathbf{y}$ given $\mathbf{x}$ while the discriminator $D$ distinguishes between real pairs $(\mathbf{x}, \mathbf{y})$ and generated pairs $(\mathbf{x}, G(\mathbf{x}, \mathbf{z}))$. Image-to-image translation encompasses both paired and unpaired approaches: in paired translation, models are trained on datasets where each source image has a corresponding target image. Conversely, unpaired translation does not rely on explicit image pairing between domains, but on learning to translate images across domains using unsupervised techniques, aiming to capture domain relationships without direct correspondences.

~\citet{DBLP:conf/cvpr/IsolaZZE17} proposed Pix2Pix, the first notable paired translation model employing a cGAN architecture with one generator and one discriminator. The generator learns to translate input images to the desired output domain conditioned on input images, while the discriminator evaluates the realism of the generated images. In contrast, CycleGAN, proposed by \citet{DBLP:conf/iccv/ZhuPIE17}, is an unpaired translation model consisting of four neural networks trained end-to-end. It employs two generator-discriminator pairs to translate images between domains. The generators learn to transform images, while the discriminators assess the authenticity of the generated images. CycleGAN leverages cycle consistency loss to enforce bidirectional translation without paired data. Typically, unpaired image-to-image translation models exhibit superior generative capabilities since the generator learns to create new content without any supervision signal, relying solely on the signal derived from cycle consistency loss. However, these models are often more difficult to train due to the increased number of neural networks involved in the end-to-end training process, as compared to paired models. This motivates our architectural choice, which is built upon a paired model, specifically Pix2Pix, with modifications to enhance the diversity of the generated outputs. In fact, improving the quality of the generated images can increase retrieval performance~\cite{DBLP:conf/cvpr/DeldjooNMM21}, allowing to learn more details of generated images in the latent space of the generative retrieval algorithm.

While GAN models are capable of effectively tackling and solving complex tasks via adversarial learning, such objective is a double-edged sword: while it drives the generator to improve continually, it can also lead to instability. If the discriminator becomes too strong, it can easily distinguish between real and fake samples, providing little information to the generator; conversely, if the generator becomes too strong, the discriminator may fail to provide meaningful feedback~\cite{DBLP:conf/iclr/MiyatoKKY18}. This imbalance can cause the generator to focus on a narrow subset of the data distribution that consistently fools the discriminator, leading to mode collapse~\cite{DBLP:conf/ijcnn/Thanh-Tung020, DBLP:conf/icml/MeschederGN18, DBLP:conf/iclr/ArjovskyB17}. This occurs when the generator learns few modes of the original data distribution that actually fool the discriminator, thus leading the training to divergence. The original GAN loss formulation, which is based on Jensen-Shannon (JS) divergence, works well when the distributions of real and generated data have little overlap. Alternative loss functions, such as those based on the Wasserstein distance~\cite{DBLP:conf/icml/ArjovskyCB17}, have been proposed to address this issue by providing smoother gradients, which can help mitigate mode collapse to some extent. This loss leads to more stable training dynamics, but, employing the Wasserstein distance necessitates that the discriminator be Lipschitz-1 continuous. To enforce this condition, weight clipping was initially used. Subsequently, it was shown that gradient penalty~\cite{DBLP:conf/nips/GulrajaniAADC17} or spectral normalization~\cite{DBLP:conf/iclr/MiyatoKKY18} could also be effectively employed to maintain the Lipschitz constraint.

Despite these advancements, mode collapse can still occur when the task involves learning complex mappings between the source and target distributions, such as in image-to-image translation between top and bottom distributions. While initially designed for minor image modifications, it is not directly applicable to the top-bottom retrieval domain due to the distinct structural characteristics of tops and bottoms. To address this issue, we propose a strategy that enhances the diversity of generated samples while reducing the likelihood of mode collapse, as described in the next sections.

\subsection{Stable Diffusion}
 
Given the inherent limitations of adversarial training discussed earlier, researchers shifted their focus to other generative models. In particular, Diffusion Models~\cite{DBLP:conf/cvpr/RombachBLEO22,DBLP:conf/nips/HoJA20} gained importance due to the highly realistic samples that such models can generate, which outperforms GANs~\cite{DBLP:conf/nips/DhariwalN21} while retaining stable training.
Diffusion models work with the forward and the backward processes. The forward process gradually corrupts the input pattern by injecting noise according to a specific Stochastic Differential Equation (SDE) process, fixed at different timesteps. In contrast, the backward process attempts to estimate the noise injected at each timestep using a complex neural network. During this backward process, the estimated noise is progressively removed from the corrupted pattern to reconstruct the original input, thereby approximating another SDE. Recent works have proposed estimating the SDE via a~\unet architecture~\cite{DBLP:conf/miccai/RonnebergerFB15}. Additionally, various samplers can be employed for generating samples from the SDE learnt by the Diffusion Model~\cite{DBLP:conf/nips/KarrasAAL22, DBLP:conf/iclr/SongME21, DBLP:conf/nips/HoJA20}. Depending on the chosen fixed forward process SDE and the number of timesteps, several strategies have been developed to enhance performance~\cite{DBLP:conf/iclr/SongME21, DBLP:conf/iclr/0011SKKEP21, DBLP:conf/iclr/Liu0LZ22}.

Despite their high-quality generations, Diffusion Models have yet to see widespread application within the fields of information retrieval and recommender systems. This contrasts VAEs and GANs, which have been extensively explored and integrated into these domains. The primary barrier to adopting Diffusion Models is the significant computational cost associated with their architectures~\cite{DBLP:conf/cvpr/RombachBLEO22,berthelot2024estimating}. These models require powerful computational resources for training and inference due to their iterative nature, which involves multiple steps of noise injection and removal through complex neural networks, thus increasing the computational resources required compared to other generative models. As a result, despite their superior ability to generate realistic samples, the practical deployment of Diffusion Models in real-time applications remains challenging. The high computational overhead limits their integration into systems where efficiency and rapid response times are essential, such as CoIR systems, where rapid responsiveness is crucial~\cite{DBLP:conf/iccv/0002OTG21}.

%% file: src/methodology.tex
\section{Methodology}\label{sec:4}
In this section, we present the proposed compatibility modeling strategy, which comprises two distinct stages designed to facilitate fashion item retrieval through generative techniques. Specifically, our proposal can be conceptualized as a two-stage solution, leveraging the outputs of an adapted Pix2Pix GAN architecture, referred to as the Complementary Item Generation Model (CIGM): the task of the~\cigm model is to generate templates of complementary items corresponding to a given input top. It is trained to perform an image-to-image translation task, learning a mapping between top images and their corresponding bottom distributions. Recalling the structure depicted in~\cref{fig:cgcrm_Workflow}, the workflow can be summarized as follows:
\begin{enumerate}
    \item \textbf{Stage 1}: Given a top image, \textit{\cigm} generates the corresponding template image representing a compatible bottom.
    \item \textbf{Stage 2}: With the top image and its corresponding template image composing the query, the \textit{\cgcrm} model computes the compatibility scores between the query and candidate bottom garments. These scores can then be utilized to perform retrieval, enabling the ranking of a list of potential candidate bottoms.
\end{enumerate}

The two models,~\cigm and~\cgcrm, are not trained in and end-to-end manner, differently from previous works. Initially, the~\cigm is trained to optimize its objective function, focusing on generating realistic and compatible template images. Subsequently, the trained~\cigm is employed within the~\cgcrm framework for compatibility modeling. We define this methodology as a \textit{two-stage compatibility modeling} approach, where the first stage involves template generation, and the second stage entails compatibility assessment. The following sections explore the problem formulation and the details of the~\cigm and~\cgcrm models.

\subsection{Problem Formulation}

We recall that we present a methodology that addresses compatibility modeling and complementary item retrieval tasks using a generative model. To achieve this, we need to model the probability distributions over the sets of tops $\mathcal{T}$ and bottoms $\mathcal{B}$. Our approach first involves learning a mapping between the probability distributions of tops $P_{data}(\mathcal{T})$ and bottoms $P_{data}(\mathcal{B})$ using the~\cigm model. Then we use this mapping to generate meaningful templates given a top $t \in \mathcal{T}$. Ideally, the~\cigm model should learn to generate samples $b \in \mathcal{B}$. However, in practice,~\cigm learns a probability distribution $P_{\cigm}(\mathcal{B})$ that approximates the probability distribution over the bottom set, such that $P_{\cigm}(\mathcal{B}) \approx P_{data}(\mathcal{B})$. Using the generated template, we employ the~\cgcrm model to assess the top-bottom compatibility.

Formally, given a dataset, we define $\mathcal{P}$ as the set of all compatible pairs $(t_{i}, b_{j})$, with $t_{i} \in \mathcal{T}$ and $b_{j} \in \mathcal{B}$, and $\mathcal{N}$ as the set of all incompatible pairs $(t_{i}, b_{k})$, with $b_{k} \in \mathcal{B}$. It is noteworthy that we consider only the pairs reported in the training data as compatible, treating all other combinations as negative examples.
We first train the~\cigm model to generate the bottom $\hat{b}_{i}$ corresponding to the input top $t_{i}$, following~\cref{eq:G}. Once the~\cigm model is trained, we train the~\cgcrm model to learn hidden compatibility relations. This is accomplished by considering the projection onto a fixed latent space of the top $t_{i}$ and the template $\hat{b}_{i}$, which are referred to as the \textit{query}. 
The model learns to align the query latent space with the bottom one, thereby encouraging compatible elements to be more similar, as detailed in the~\Cref{eq:C}. We first explain the model used to create templates and then discuss the~\cgcrm model.

\begin{equation}
    t_{i} \xrightarrow{\text{\cigm}} \hat{b}_{i}, \quad \forall t_{i} \in \mathcal{T}
\label{eq:G}
\end{equation}

\begin{equation}
(t_{i}, \hat{b}_{i}) \xrightarrow{\text{\cgcrm}} b_{j}, \quad \forall (t_{i}, b_{j}) \in \mathcal{T}, \mathcal{B}
\label{eq:C}
\end{equation}

\begin{figure}[t]
    \centering
    \includegraphics[width=0.4\textwidth]{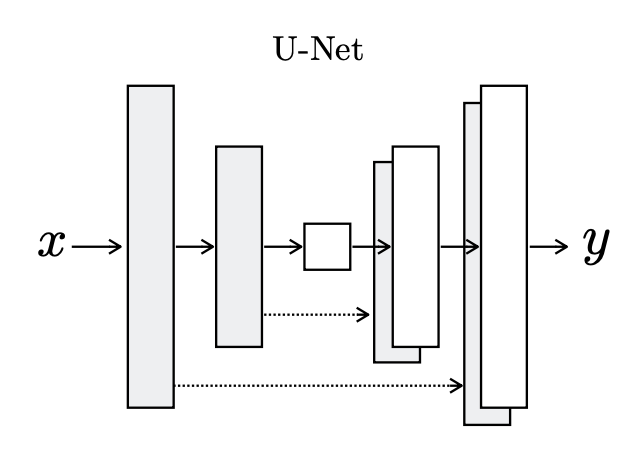}
    \caption{Pix2Pix original generator~\cite{DBLP:conf/cvpr/IsolaZZE17}.}
    \label{fig:pix2pix_generator}
\end{figure}

\subsection{Complementary Item Generation Model \textit{(CIGM)}}

The objective of \cigm is to learn a mapping between the $P_{data}(\mathcal{T})$ and $P_{data}(\mathcal{B})$, computing the $P_{\cigm}(\mathcal{B})$ probability distribution, which should be as close as possible to $P_{data}(\mathcal{B})$. This leads to the adoption of an image-to-image translation architecture to accomplish the conditioning generation task, to learn a mapping between the source and target distributions.

Specifically, our generative model is based on the Pix2Pix~\cite{DBLP:conf/cvpr/IsolaZZE17} architecture, which consists of two primary components: a \unet generator, depicted in \cref{fig:pix2pix_generator}, and a PatchGAN discriminator. The \unet generator works as an autoencoder with skip-connections, which help to address the vanishing gradient problem~\cite{DBLP:conf/icml/PascanuMB13} during training. Conversely, the PatchGAN discriminator generates a patch, which is a specific portion of the image of a pre-defined size, rather than producing a single scalar output. Ideally, this patch contains values close to zero or one, indicating the level of realism of the patch. The primary objective is to leverage local \textit{receptive fields}~\cite{DBLP:conf/cvpr/IsolaZZE17}, facilitating the capture of high-frequency details and textures with greater effectiveness. This design allows the discriminator to identify which specific areas of the generated image should be improved to deceive the discriminator effectively.

\begin{figure}[t]
    \centering
    \includegraphics[width=0.9\textwidth]{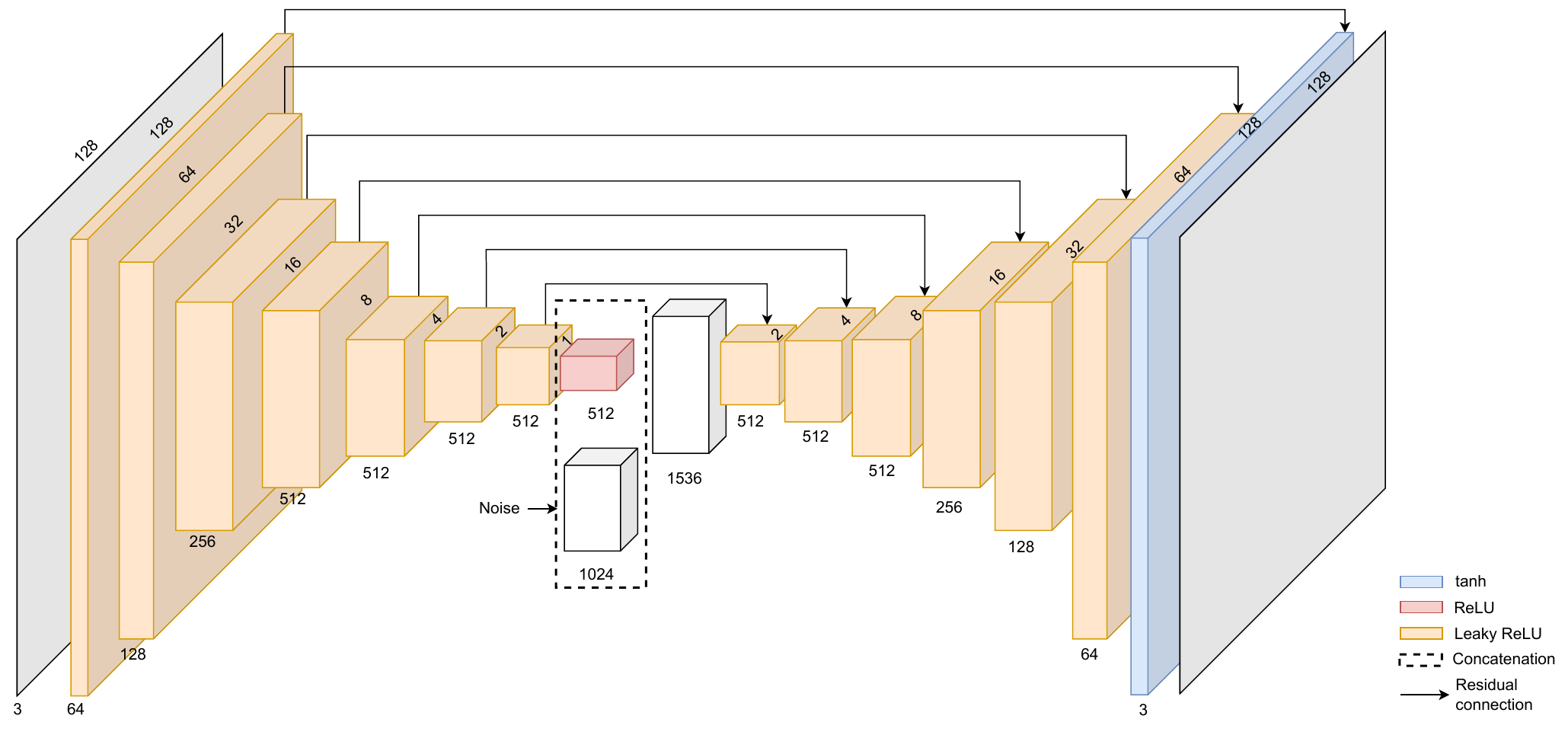}
    \caption{Complementary Item Generation Model.}
    \label{fig:custom_pix2pix_generator}
\end{figure}

As foreshadowed in the previous section, in order to overcome the mode collapse phenomenon~\cite{DBLP:conf/ijcnn/Thanh-Tung020,DBLP:conf/icml/MeschederGN18} and to produce more realistic templates, we modified the generator architecture to incorporate noise at the bottleneck level. Specifically, the latent representation is concatenated with a noise vector $\mathbf{z} \sim p(\mathbf{z})$, where $p(\mathbf{z})$ is a Gaussian distribution. This noise injection aims to add stochasticity to the generation process, thereby increasing the diversity of the generated samples, while maintaining the simplicity inherent to a paired image-to-image translation model. Additionally, the skip connections are retained to leverage the benefits of the \unet architecture for better feature propagation during training. \Cref{fig:custom_pix2pix_generator} illustrates the adapted architecture.

The CIGM model is trained considering two different loss components:

\begin{itemize}
    \item \textit{Adversarial loss}: This component derives from the JS divergence and evaluates the Binary Cross-Entropy (BCE) between the model's output and the target labels, which are determined by the Patch Discriminator's evaluation of the generated samples.
    \item \textit{L1 loss}: This component evaluates the L1 norm of the difference between the generated templates and the corresponding ground-truth bottoms.
\end{itemize}

Formally, we define the generator as shown in~\Cref{eq:G}, and the discriminator following~\Cref{eq:d}, where $c$ refers to the dimension of the patch.

\begin{equation}
    (t_{i}, b_{j}) \xrightarrow{D} 1^{c \times c}, (t_{i}, \hat{b}_{i}) \xrightarrow{D} 0^{c \times c}, \quad \forall (t_{i}, b_{j}) \in \mathcal{T}, \mathcal{B}
    \label{eq:d}
\end{equation}
The objective functions are defined as:
\begin{equation}
\min_G \max_D \mathcal{L}_{\text{GAN}} = \mathbb{E}_{t_i \sim \mathcal{T}, b_j \sim \mathcal{B}} [\log D(t_i, b_j)] + \mathbb{E}_{t_{i} \sim \mathcal{T}, z \sim p_{z}(z)} [\log (1 - D(t_i, G(t_i, z)))]
\label{eq:cost_minmax}
\end{equation}

\begin{equation}
\mathcal{L}_G = -\mathbb{E}_{t_{i} \sim \mathcal{T}, z \sim p_{z}(z)} [\log D(t_i, G(t_{i}, z))] + \lambda \| G(t_{i}, z) - b_{j} \|_{1}
\label{eq:cost_g}
\end{equation}

\begin{equation}
\mathcal{L}_D = -\mathbb{E}_{t_i \sim \mathcal{T}, b_j \sim \mathcal{B}} [\log D(t_i, b_j)] - \mathbb{E}_{t_i \sim \mathcal{T}, z \sim p_{z}(z)} [\log (1 - D(t_i, G(t_i, z)))]
\label{eq:cost_d}
\end{equation}

\begin{figure}[b]
    \centering
    \begin{subfigure}[b]{\textwidth}
        \includegraphics[width=\textwidth]{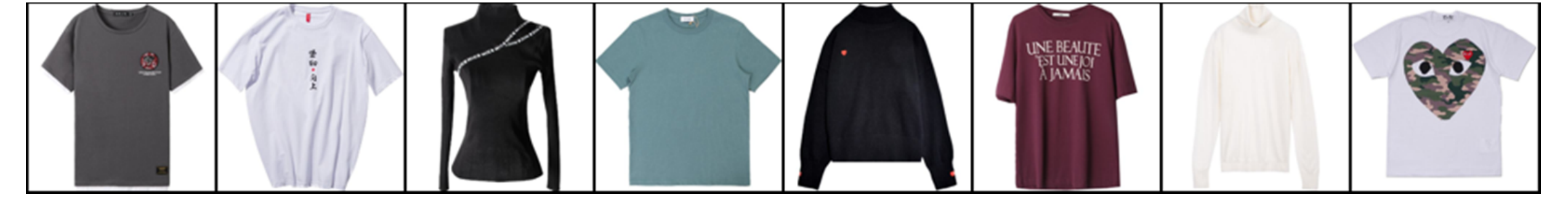}
        \caption{Conditioning tops.}
        \label{subfig:noisy_unet_input}
    \end{subfigure}
    \hfill
    \begin{subfigure}[b]{\textwidth}
        \includegraphics[width=\textwidth]{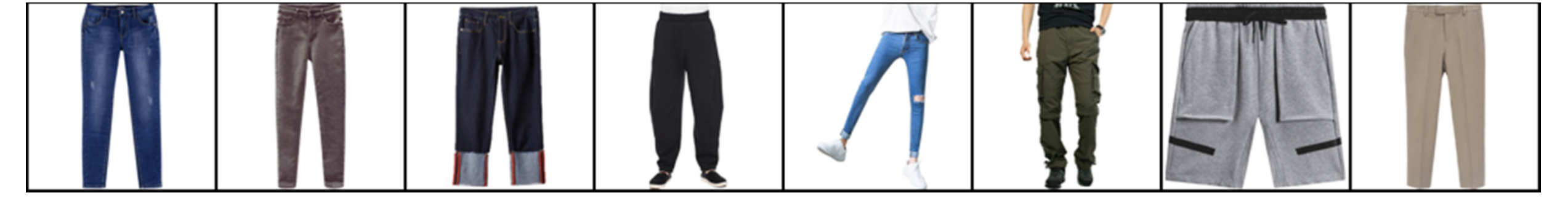}
        \caption{Ground-truth bottoms.}
        \label{subfig:noisy_unet_label}
    \end{subfigure}
    \hfill
    \begin{subfigure}[b]{\textwidth}
        \includegraphics[width=\textwidth]{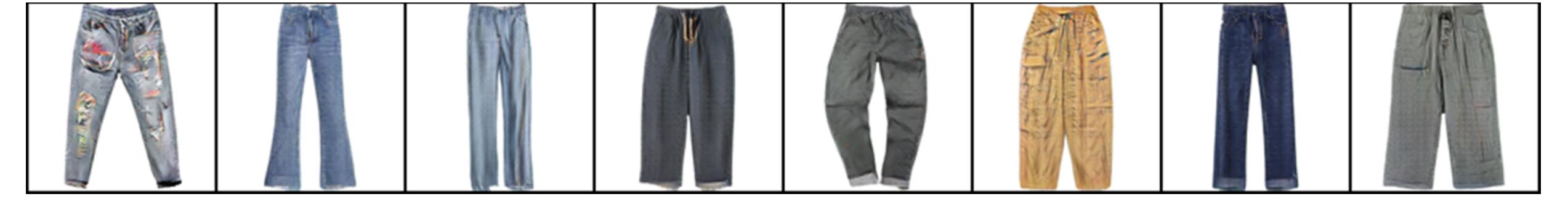}
        \caption{Generated bottoms with the proposed Generative Model.}
        \label{subfig:noisy_unet_gen}
    \end{subfigure}
    \caption{Top: conditioning tops. Middle: ground-truth bottoms. Bottom: generated bottoms with the proposed generative model.}
    \label{fig:noisy_unet_generations}
\end{figure}
Given the min-max game expressed in \cref{eq:cost_minmax}, the generator aims to minimize the objective in \cref{eq:cost_g}, where $\| G(t_{i}, \mathbf{z}) - b_{j} \|_{1}$ is a regularization term that guides the generator, weighted by the hyperparameter $\lambda$. Conversely, the discriminator aims to minimize the objective in \cref{eq:cost_d}. The goal is to reach a Nash equilibrium where the generator successfully fools the discriminator. Once this equilibrium is achieved, only the generator is employed in the proposed \textit{\cgcrm} model. The discriminator's role is solely to guide the generator in learning the underlying distribution. \cref{fig:noisy_unet_generations} displays some samples using the modified \unet generator with added noise. As depicted in~\cref{subfig:noisy_unet_gen}, the generated bottom images are not only realistic but also exhibit diversity compared to the actual ground-truth shown in~\cref{subfig:noisy_unet_label}.

\subsection{Generative Compatibility Model \textit{(\cgcrm)}}
\begin{figure}[t]
    \centering
    \includegraphics[width=\textwidth]{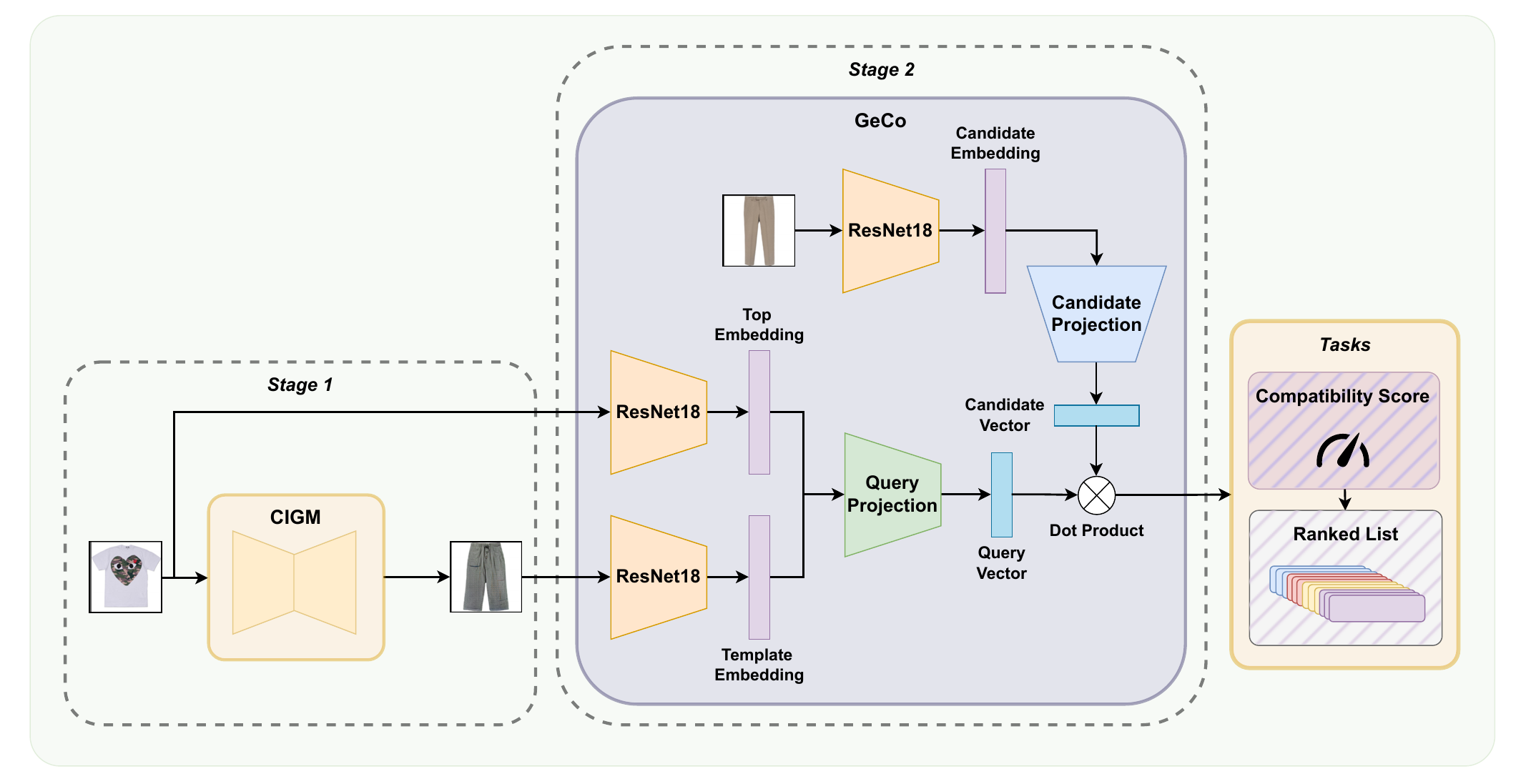}
    \caption{The complete \textit{two-stage} architecture, highlighting the Generative Compatibility Model (\textit{\cgcrm}) and describing the overall structure.}
    \label{fig:cgcrm}
\end{figure}

The~\cgcrm model can create meaningful queries by harnessing realistic templates generated by the~\cigm model. In contrast to previous approaches,~\cgcrm integrates the templates into the query~\cite{DBLP:journals/ijon/LiuSCM20,DBLP:conf/ijcai/LiuSRNT020,DBLP:conf/www/LinRCRMR19}, thereby adhering to a CoIR task framework. As discussed before, this method puts more emphasis to the generated templates, rather than treating them merely as regularization signals. Consequently, we consider only the visual signal to evaluate the effectiveness of our methodology.

The objective of the learning procedure is to better align the query and bottom embedding space better, thereby maximizing the compatibility between the fashion items. Specifically, 
\cgcrm employs a pre-trained ResNet18~\cite{DBLP:conf/cvpr/HeZRS16} backbone to project the triplet $(t_{i}, \hat{b}_{i}, b_{j})$ onto a feature space $F \subset \mathbb{R}^{512}$. Subsequently, the latent representations $\tilde{t_{i}}, \tilde{b_{i}}$ are concatenated and projected via a linear layer onto a lower-dimensional space $W \subset \mathbb{R}^{128}$. Similarly, $\tilde{b_{j}}$ is projected onto another space $\tilde{W} \subset \mathbb{R}^{128}$, with the objective of aligning the spaces $W$ and $\tilde{W}$.~\cref{fig:cgcrm} shows the overall structure of the~\textit{\cgcrm} model.

We argue that maximizing the compatibility is equivalent to maximize the similarity between vectors lying in the two spaces $W$ and $\tilde{W}$, measured via a simple dot product. To achieve this objective, we have formulated the loss function taking into account both BPR~\cite{DBLP:conf/uai/RendleFGS09} and Contrastive Learning objectives, namely the Information Noise-Contrastive Estimation (\nce)~\cite{DBLP:journals/corr/abs-1807-03748} loss.
While the BPR loss focuses on optimizing pairwise rankings by ensuring preferred items are ranked higher than non-preferred items, the \nce loss operates in a self-supervised manner, aiming to bring positive sample pairs closer together in the latent space while pushing negative sample pairs farther apart.
Thus, optimizing these two loss functions maximizes the similarity among compatible pairs while considering all the other bottoms as negatives.

Formally, given a set $\mathcal{Q} \coloneqq \{(i, j, k) \mid (t_{i}, b_{j}) \in \mathcal{P}, \quad b_{k} \in \mathcal{P}\setminus b_{j}\}$, we define the compatibility score between the positive top-bottom pairs and the negative ones in the \cref{eq:mijk}, where $m_{ij}$ represents the positive score, $m_{ik}$ represents the negative scores, and $\text{sim}(\cdot, \cdot)$ denotes the dot product.

\begin{equation}
    \begin{aligned}
        m_{ij} = \text{sim}(t_{i}, b_{j}), \\
        m_{ik} = \text{sim}(t_{i}, b_{k})
\end{aligned}\label{eq:mijk}
\end{equation}

We define the complete loss in \cref{eq:whole_cgcrm_loss}, where the strength of each component in the overall loss $\mathcal{L}$ is regulated by the hyperparameters $\alpha$, $\beta$, and $\gamma$.

\begin{equation}
    \mathcal{L} = \alpha \cdot \mathcal{L}_{\text{bpr}} + \beta \cdot \mathcal{L}_{\text{\nce}} + \gamma \cdot \mathcal{L}_{\text{reg}}
    \label{eq:whole_cgcrm_loss}
\end{equation}

We define the BPR loss in \cref{eq:bpr} and the~\nce loss in \cref{eq:nce}, where $\tau$ is the temperature parameter, and the regularization component in \cref{eq:reg} with $\theta_{c}$ representing the parameters of the \cgcrm model.

\begin{equation}
    \mathcal{L}_{\text{bpr}} = -\log \sigma (m_{ij} - m_{ik})
    \label{eq:bpr}
\end{equation}
\begin{equation}
    \mathcal{L}_{\text{\nce}} = -\frac{1}{\tau} \log \frac{\exp(\frac{1}{\tau}m_{ij})}{\sum_{k=1}^{K}\exp(\frac{1}{\tau}m_{ik})}
    \label{eq:nce}
\end{equation}
\begin{equation}
    \mathcal{L}_{\text{reg}} = \|\theta_{c}\|_{\scriptscriptstyle 2}
    \label{eq:reg}
\end{equation} 

%% file: src/experiments.tex
\section{Experiments}\label{sec:5}
Aiming to demonstrate the effectiveness of our model, we conducted a benchmarking study using three different datasets and compared the results with various baselines. Our primary objective is to determine whether the proposed \textit{two-stage} model effectively achieves the desired outcomes, as measured by the selected metrics, compared to other benchmarks.

We first detail the selected datasets. Afterward, we outline the metrics utilized in our evaluation. Additionally, we discuss implementation details. Lastly, we present and analyze the achieved results, incorporating an ablation study and the impact of individual hyperparameters. This analysis aims to showcase the effectiveness of the combined loss function compared to utilizing solely the BPR or~\nce components.

\subsection{Datasets}

To evaluate the effectiveness of \cgcrm, we conducted experiments using three distinct datasets: (i) FashionVC~\cite{DBLP:conf/mm/SongFLLNM17}, (ii) ExpFashion~\cite{DBLP:journals/tkde/LinRCRMR20}, and (iii) FashionTaobaoTB. The FashionVC dataset comprises 20,726 outfits, including 14,870 tops and 13,662 bottoms. ExpFashion contains 853,991 outfits, with 168,682 tops and 117,668 bottoms. Both FashionVC and ExpFashion datasets were sourced from the fashion-sharing platform Polyvore. To facilitate a more direct comparison with the baselines, we followed the approach in~\cite{DBLP:journals/ijon/LiuSCM20}. We created a reduced version of ExpFashion, termed ExpReduced, which matches the number of outfits in FashionVC. The resulting ExpReduced dataset includes 17,140 positive top-bottom combinations, sampled randomly with a fixed random seed.

\input{tables/datasets_characteristics}

In addition, \citet{DBLP:conf/kdd/ChenHXGGSLPZZ19} introduced a new dataset from the e-commerce site Taobao, consisting of 1,000,000 outfits with over 5,000,000 items. This dataset includes various outfit types composed of different fashion items. For our problem setting, we curated a subset named FashionTaobaoTB. This new dataset was meticulously crafted through data cleaning and visual inspection to select relevant items, given the lack of clear mapping between category IDs and their corresponding classes (e.g., top, bottom, shoes). To the best of our knowledge, this study represents the first adaptation of this dataset for top-bottom retrieval.

The characteristics of the three training datasets are summarized in \cref{table:dataset_characteristics}. Notably, the distribution of tops and bottoms in each dataset highlights that the FashionTaobaoTB dataset contains a larger number of tops and bottoms, making it a valuable resource for benchmarking in this domain.

\subsection{Evaluation metrics}
We evaluate our models using two metrics, which were previously used in the literature for top-bottom compatibility modeling: Area Under the Curve (AUC)~\cite{DBLP:journals/prl/Fawcett06} for compatibility modeling and Mean Reciprocal Rank (MRR)~\cite{DBLP:journals/computer/KorenBV09} for complementary item retrieval, based on~\cref{eq:mijk}. We recall that $m_{ij}$ denotes the compatibility score between the positive top-bottom pairs, whereas $m_{ik}$ represents the compatibility score between the negative top-bottom pairs.

The AUC is defined below, where $\delta(\cdot, \cdot)$ is 1 if $m_{ij} > m_{ik}$, otherwise 0. We recall that $m_{ij}$ and $m_{ik}$ convey the similarity score between a pair of top $t_i$ and bottom $b_j$ or $b_k$ as formulated in~\cref{eq:mijk}.
\begin{equation}
    \textit{AUC} = \frac{1}{|\mathcal{P}|} \sum_{i,j \in \mathcal{P}} \delta(m_{ij}, m_{ik})
    \label{eq:auc}
\end{equation}
The MRR is defined in~\cref{eq:mrr}, where $\text{rank}_i$ refers to the position of the positive element within the ranked list.
\begin{equation}
    \textit{MRR} = \frac{1}{|\mathcal{P}|} \sum_{i=1}^{|\mathcal{P}|} \frac{1}{\text{rank}_i + 1}
    \label{eq:mrr}
\end{equation}
We compute the AUC by evaluating each top $t_i$ with its corresponding positive bottom $b_j$ and negative bottom $b_k$. It is noteworthy that in a prior work on compatibility modeling,~\citet{DBLP:conf/mm/SongFLLNM17} utilized three random negatives for AUC and nine random negatives for MRR. Unfortunately, there is no clear reference for the methodology by which these negatives were sampled or justifying this choice, which is why we compute MRR by considering the rank position of the positive bottom $b_j$ within the ranked list, where all other bottoms in the test set are treated as negatives.

\subsection{Implementation Details}

We re-implemented the baseline models from the original paper descriptions due to outdated codebases and library incompatibilities with our system architecture. This re-implementation enabled us to execute the new codebase on our server, which is equipped with a single Nvidia A10 GPU, an AMD EPYC-Rome CPU, and 64GB of RAM. We used the PyTorch framework for all model implementations, taking advantage of its flexibility and efficiency for deep learning tasks. The hyperparameters for the re-implemented models were aligned with those specified in the original papers for the FashionVC and ExpReduced datasets. 
For the FashionTaobaoTB dataset, we explored the hyperparameters within the same ranges as those used in the original experiments.
Additionally, all image data was consistently resized to dimensions of $128 \times 128$.

The \textit{\cigm} model was trained for 200 epochs with a learning rate of $2 \times 10^{-4}$, a batch size of 64, and a regularization parameter ($\lambda$) set to 100. For the \textit{\cgcrm} model, a learning rate of $1 \times 10^{-4}$ was utilized and decayed by a factor of 0.1 every 8 epochs using the StepLR scheduler throughout 50 epochs, with a batch size of 64. The backbone architecture was a pre-trained ResNet18, fine-tuned end-to-end alongside the projection layer. The projection layer consisted of two simple dense layers, facilitating projection into a 128-dimensional space. We found that employing more complex architectures for the projection layer led to a decrease in performance.

\subsection{Baseline Models}
Aiming to assess the efficacy of~\cgcrm, we compare our model against the following baseline methods. We recall that this evaluation focuses solely on the processing of the visual signal, as discussed in the previous sections.

\begin{itemize}
    \item \textbf{Random}: A baseline model that assigns random compatibility scores between tops and bottoms.
    
    \item \textbf{Popularity}: the compatibility score between top and bottom is based on ``popularity'', quantified as the number of tops associated with the bottom \( b_j \), following~\cite{DBLP:conf/mm/SongFLLNM17}.

    \item \textbf{BPR-DAE}~\cite{DBLP:conf/mm/SongFLLNM17}: a content-based clothing matching scheme that jointly models the compatible preferences of fashion items with multi-modalities and the coherent relation among different modalities of the same item. We adapt this method to encode only the visual features using the original AlexNet backbone, as discussed and evaluated in~\cite{DBLP:journals/ijon/LiuSCM20}.

    \item \textbf{MGCM}~\cite{DBLP:journals/ijon/LiuSCM20}: a compatibility schema based on a GAN model that computes fashion item compatibility by considering both item-item and item-template compatibility scores. The item-item score represents the actual compatibility between the top and the candidate bottom embeddings, while the item-template score is the \(L_1\) norm between the candidate bottom and the generated embeddings.

    \item \textbf{Pix2PixCM}~\cite{DBLP:journals/ijon/LiuSCM20}: Similar to MGCM, but with the generator replaced by a Pix2Pix GAN model.

\end{itemize}

\input{tables/1vs1_disj}

\cref{table:1vs1} presents the results obtained using the proposed evaluation protocol. It is evident that \cgcrm outperforms all other models across all datasets.
Notably, BPR-DAE emerges as the second-best performer, exhibiting a significant margin over other models on the FashionVC and ExpReduced datasets. Conversely, MGCM achieves AUC values comparable to BPR-DAE only on the FashionTaobaoTB dataset, albeit still lower than \cgcrm, while exhibiting the second-highest MRR. This observation suggests that training an end-to-end generative compatibility model, such as MGCM, delivers high performance mainly when the training data contains more information. The same holds for the other generative baseline used, namely Pix2PixCM. Conversely, on smaller datasets such as \textit{FashionVC} and \textit{ExpReduced}, only simpler strategies like BPR-DAE demonstrate competitive performance. Other generative models struggle to achieve satisfactory results, thereby highlighting the efficacy of our model on these datasets. Considering Popularity and Random baselines, we observe a not common behavior. Indeed, these results reveal that Random performs superior to Popularity: this is related to the distributions of pairs in the datasets analyzed. As shown in~\Cref{fig:pop_dist}, most bottoms are paired with only a limited number of tops, indicating that these bottoms are not really popular, as they tend to be recommended without a clear underlying rationale. Consequently, the Popularity baseline assigns relatively uniform compatibility scores across numerous items, which reduces its effectiveness and affects its performance in comparison to the Random baseline.

We note that the baseline results reported in \cref{table:1vs1} differ from the results in~\cite{DBLP:journals/ijon/LiuSCM20} due to differences in how the metrics are computed. As mentioned above,~\citet{DBLP:journals/ijon/LiuSCM20} calculated the AUC and MRR metrics in an uncommon approach by pairing each positive item with three and nine negative items, respectively. To understand how our proposed model would fare in such a setup, we repeated the measurement using the specific MRR and AUC calculation from~\cite{DBLP:journals/ijon/LiuSCM20}, leading to the results shown in \cref{table:1vs3}. We also find that ~\cgcrm outperforms all considered baselines under this uncommon configuration. BPR-DAE emerges as the second-best model in most cases, except for the FashionTaobaoTB dataset, where MGCM benefits from a larger volume of data, enabling it to learn more meaningful relations. Once more, our model demonstrates its adaptability to varying sample sizes in the training set. Furthermore, Popularity and Random retain the same behavior as the previous evaluation setting.

\input{tables/1vs3_disj}

\input{figures/distribution}

\subsection{Ablation Study}

In the previous sections, we demonstrated that our model can accurately pair top-bottom clothing items and generate highly realistic templates. Furthermore, the \cgcrm model performs
better than existing models for the ranking task across all considered datasets.

The dual nature of our model, addressing compatibility and ranking, stems from our choice of the loss function. \Cref{eq:whole_cgcrm_loss} presents the loss function as a weighted sum of two components and a regularization term. We focus on the contributions from the two main components—BPR and \nce by evaluating how these components affect the final results and whether both are necessary to achieve the model's promised performance is essential.

The objective of this ablation study is to understand the contributions of the two main components of the loss function to the final performance of the model. The BPR and \nce components are weighted by parameters $\alpha$ and $\beta$, respectively. We consider three versions of the loss function in our ablation study: i) only the BPR component active ($\beta = 0$), ii) only the \nce component active ($\alpha = 0$), and iii) both components active with weighted contributions ($\alpha \neq 0$ and $\beta \neq 0$).

\Cref{table:ablation_study} presents the results of this comparative study of different contributions to the loss function. We observe that the weighted sum of the two components offers the best trade-off, consistently providing a high AUC and excellent ranking performance. Notably, with only the~\nce component active, there is a significant improvement in the ranking metric, as expected, while the AUC degrades markedly with only the BPR component active. This behavior is due to two factors. First, our model is optimized for the AUC metric using a weighted loss ($\alpha$ and $\beta$ not equal to zero), making this condition the most promising for performance. Second, the \nce component positively contributes to ranking and compatibility performance~\cite{DBLP:conf/aaai/HoffmannBGBN22}. Furthermore, the \nce loss demonstrates robust performance across all datasets, including those with a limited number of training samples, such as FashionVC and ExpReduced. In the next subsection, we dive deeper into the effects on model performance by exploring the parameters $\alpha$, $\beta$, and $\tau$.

\input{tables/ablation}

\input{figures/alpha}

\subsubsection{\textbf{Weight Variation Analysis}}

Aiming to analyze how to appropriately combine the parameters of the loss function presented in \cref{eq:whole_cgcrm_loss},
we conducted comprehensive hyperparameter tuning across various datasets to examine the sensitivity and significance of the loss parameters in our model. 
Specifically, for $\alpha$ and $\beta$ we explored the range $[0.25, 0.5, 0.75]$; for the scalar $\tau$, values in the range $[0.005, 0.01, 0.05, 0.1, 0.5]$ were considered.

Figure~\ref{fig:hp_sensitivity} presents scatter plots illustrating the changes in MRR and AUC on the FashionTaobaoTB dataset resulting from different configurations of these loss parameters. Here, $\alpha$ determines the weight of the BPR component,
$\beta$ controls the weight of the \nce component, and $\tau$ is the temperature parameter for the softmax distribution in the \nce loss, affecting the sharpness of similarity scaling. Lower $\tau$ values enhance discriminative power, while higher values soften it~\cite{DBLP:conf/cvpr/WangL21a}.
Analyzing $\alpha$, as shown in~\cref{fig:hp_alpha}, we observed that a value of 0.5 yields optimal AUC performance and maximized MRR, thus providing the best results for the complementary item retrieval task. Higher $\alpha$ values significantly decrease the MRR, indicating a shift in the loss contribution towards compatibility modeling, with $\beta=1$ and $\tau=0.5$.
~\Cref{fig:hp_beta} illustrates the effects of varying $\beta$ on model performance, setting $\alpha=0.5$ and $\tau=0.5$. Increasing $\beta$ improved both the AUC and the MRR, highlighting the significant impact of the contrastive loss on these tasks. Conversely, lower $\beta$ values greatly reduced the MRR, underscoring the critical role of this component in retrieval tasks.
The temperature parameter $\tau$ was crucial for model performance, as depicted in~\cref{fig:hp_tau}, where $\alpha=0.5$ and $\beta=1$. Setting $\tau$ to 0.1 yields the highest MRR, favoring higher softmax temperatures. The optimal AUC was achieved with $\tau$ set to 0.5, indicating the importance of temperature in the softmax for compatibility modeling and item retrieval. Lower $\tau$ values impaired the model's ability to capture the compatibility relations for these tasks accurately.

Overall, this analysis underscores the necessity of finely tuned hyperparameters, particularly regarding the interplay between the BPR and~\nce components, which significantly affect both tasks. Additionally, the $\tau$ parameter is pivotal in achieving optimal performance. The trends shown in~\cref{fig:hp_sensitivity} were consistently observed across other datasets, supporting our conclusions. In particular, for the FashionVC dataset, the best results are achieved with $\alpha=0.25$, $\beta=0.25$, and $\tau=0.5$, while for the ExpReduced dataset, the optimal combination is $\alpha=0.25$, $\beta=0.25$, and $\tau=0.5$. This analysis primarily focuses on the combination of loss parameters that optimize compatibility modeling, as measured by the AUC. Despite this, it is certainly possible to identify other combinations of these parameters that can push the ranking performance of our model at the expense of accuracy performance (e.g., as is the case with $\tau$), but this exploration and optimization is outside the scope of our work.

\subsection{Comparison of Templates}
\begin{figure}[t!]
    \centering
    \includegraphics[width=0.65\textwidth]{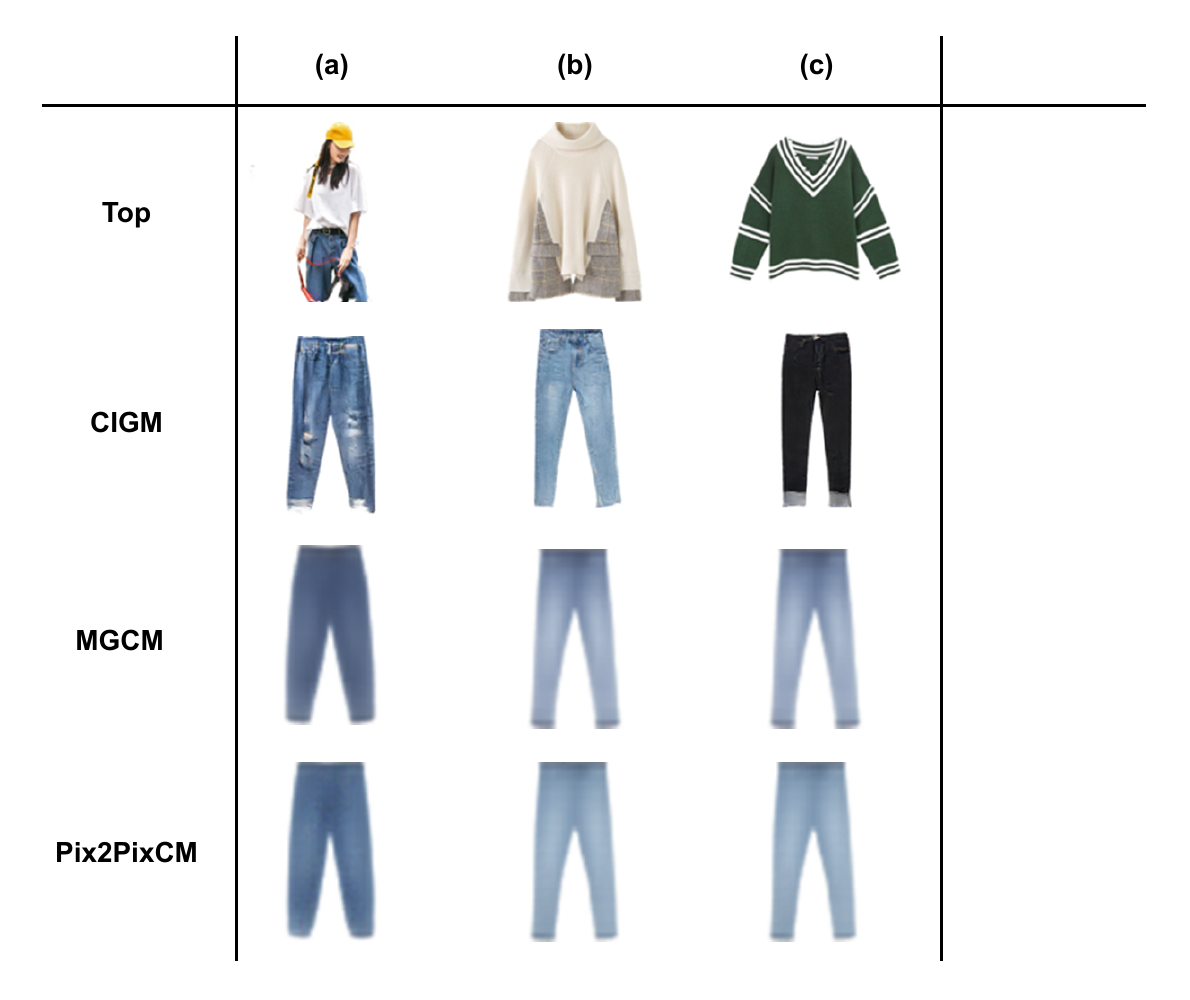}
    \caption{The templates generated by the \cigm, MGCM, and Pix2PixCM models, given the same top image, highlighting the superior quality of our templates. The images are taken from the FashionTaobaoTB dataset. We note that the templates from the baseline models (MGCM and Pix2PixCM) are scaled to match the higher resolution of our templates.}
    \label{fig:templates}
\end{figure}

Aiming to demonstrate the highest quality of the templates generated by the \cigm model utilized in \cgcrm, we present several visual examples randomly selected from the FashionTaobaoTB dataset compared to the baselines. \Cref{fig:templates} displays the generated bottom images corresponding to the three top image inputs. Evidently, the templates produced by \cigm exhibit significantly more detail than those generated by the two baseline models, which appear blurred and retain only basic color and shape information.

\begin{figure}[h!]
    \centering
    \begin{adjustbox}{width=1\textwidth}
    \includegraphics{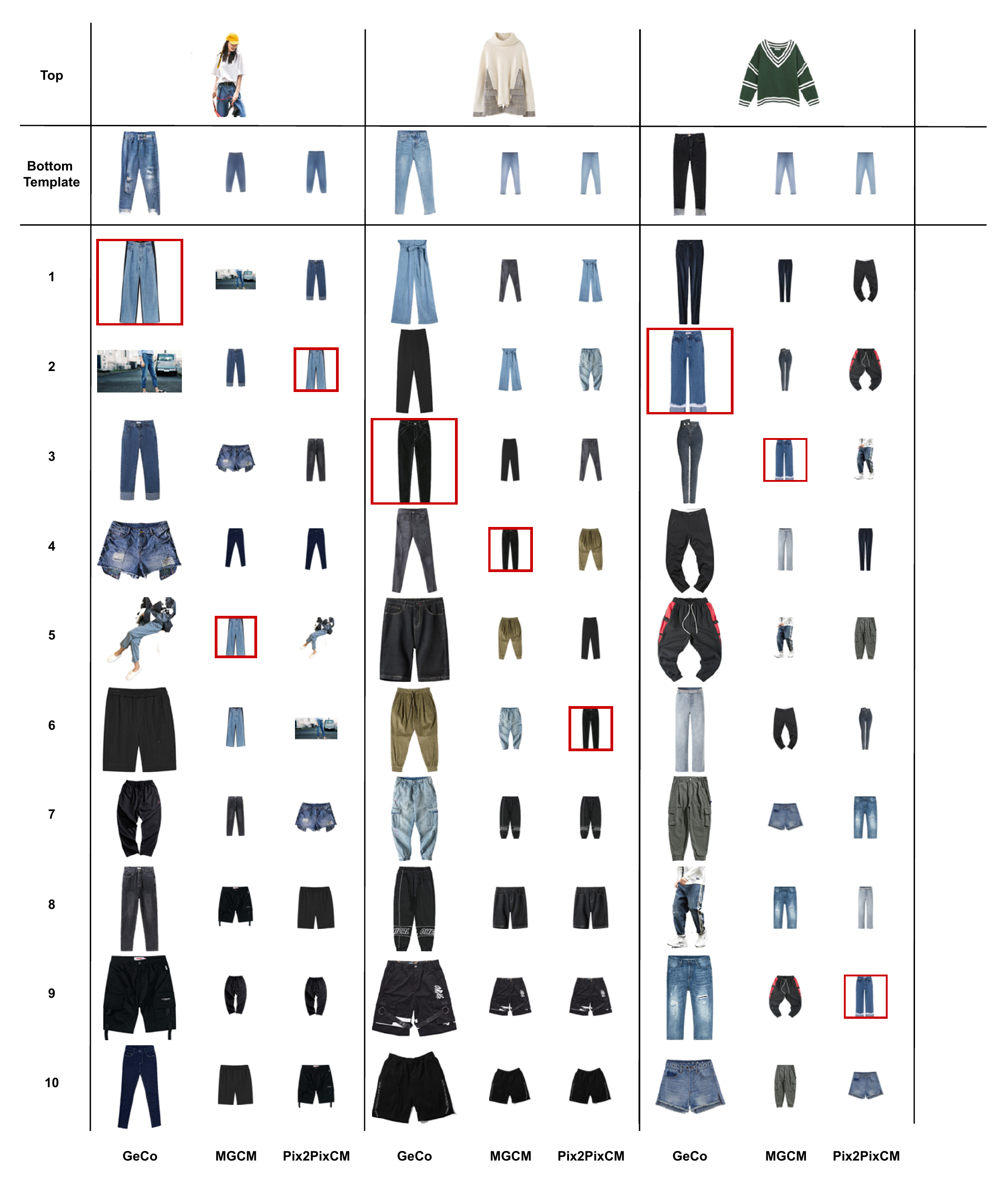}
    \end{adjustbox}
    \caption{Retrieval performance of the \cgcrm model on the FashionTaobaoTB dataset, compared to various baseline models, all using the same input top. It can be observed that our model exhibits superior retrieval performance and generates more realistic templates with higher resolution. The first row displays the input top, while the second row shows the bottom template generated by the corresponding model in each column. The following rows indicate the position of the retrieved items. The positive item is highlighted by a red box.}
    \label{fig:retrieval}
\end{figure}

Furthermore, we assess the impact of the randomly selected templates in the complementary item retrieval task, with results shown in \Cref{fig:retrieval}. The \cigm model generates templates of markedly higher quality compared to baseline methods, ensuring a four-fold resolution improvement. As highlighted by the red boxes in the figure, our model consistently outperforms the baselines by correctly identifying the positive bottom in the ranked list across all evaluated cases. Notably, integrating these templates into the retrieval pipeline sometimes leads to generated details that differ from those of the positive items while preserving compatibility information. For instance, despite the visual differences in some generated templates, the retrieved items remain visually compatible with the input top. This demonstrates the capability of our approach to enhance diversity in complementary retrieval results.

%% file: tables/datasets_characteristics.tex
\begin{table}
\centering
\caption{Summary of dataset characteristics: each of the columns show the number of instances of tops, bottoms, and positive pairs, respectively.}
\label{table:dataset_characteristics}
\begin{tblr}{
  cell{2}{2} = {c},
  cell{2}{3} = {c},
  cell{2}{4} = {c},
  cell{3}{2} = {c},
  cell{3}{3} = {c},
  cell{3}{4} = {c},
  cell{4}{2} = {c},
  cell{4}{3} = {c},
  cell{4}{4} = {c},
  hline{1-2,5} = {-}{},
}
\textbf{Dataset}                & \textbf{Tops} & \textbf{Bottoms} & \textbf{Positives} \\
FashionVC~                      & 13,809        & 12,756           & 18,640             \\
ExpReduced~                     & 12,791        & 11,026           & 18,640             \\
FashionTaobaoTB\textit{ (ours)} & 43,096        & 35,414           & 89,951             
\end{tblr}
\end{table}

%% file: tables/1vs1_disj.tex
\begin{table}[t]
\centering
\caption{Performance of the different models on three datasets. For each metric and dataset, \textbf{boldface} and \uline{underlined} numbers indicate best and second-to-best values.}
\label{table:1vs1}
\begin{tblr}{
  row{1} = {c},
  column{3} = {c},
  column{5} = {c},
  column{7} = {c},
  cell{1}{2} = {c=2}{},
  cell{1}{4} = {c=2}{},
  cell{1}{6} = {c=2}{},
  cell{2}{2} = {c},
  cell{2}{4} = {c},
  cell{2}{6} = {c},
  cell{3}{2} = {c},
  cell{3}{4} = {c},
  cell{3}{6} = {c},
  cell{4}{2} = {c},
  cell{4}{4} = {c},
  cell{4}{6} = {c},
  cell{5}{2} = {c},
  cell{5}{4} = {c},
  cell{5}{6} = {c},
  cell{6}{2} = {c},
  cell{6}{4} = {c},
  cell{6}{6} = {c},
  cell{7}{2} = {c},
  cell{7}{4} = {c},
  cell{7}{6} = {c},
  cell{8}{2} = {c},
  cell{8}{4} = {c},
  cell{8}{6} = {c},
  hline{1,3,9} = {-}{},
  hline{2} = {2-7}{},
}
\textbf{Model} & \textbf{FashionVC} &                 & \textbf{ExpReduced} &                 & \textbf{FashionTaobaoTB} &                 \\
               & \textit{AUC}       & \textit{MRR}    & \textit{AUC}        & \textit{MRR}    & \textit{AUC}              & \textit{MRR}    \\
\textbf{\cgcrm} & \textbf{0.7621}    & \textbf{0.0307} & \textbf{0.7295}     & \textbf{0.0151} & \textbf{0.8491}           & \textbf{0.0126} \\
MGCM           & ~ 0.5939~~         & ~ 0.0098~~      & ~ 0.4674~~          & ~ 0.0075~~      & 0.7508                    & \uline{0.0095}  \\
Pix2PixCM      & 0.5000             & 0.0080          & 0.5136              & 0.0078          & 0.6934                    & 0.0056          \\
BPR-DAE        & \uline{0.6590}     & \uline{0.0200}  & \uline{0.6457}      & \uline{0.0130}  & \uline{0.7531}            & 0.0052          \\
Popularity            & 0.3518             & 0.0042          & 0.3887              & 0.0020          & 0.4638                    & 0.0029                \\
Random         & 0.4908             & 0.0081          & 0.4908              & 0.0063          & 0.5046                    & 0.0014          
\end{tblr}
\end{table}

%% file: tables/1vs3_disj.tex
\begin{table}[t]
\centering
\caption{Performance of the different models on three datasets following the MGCM evaluation protocol~\cite{DBLP:journals/ijon/LiuSCM20}. For each metric and dataset, \textbf{boldface} and \uline{underlined} numbers indicate best and second-to-best values.}
\label{table:1vs3}
\begin{tblr}{
  row{1} = {c},
  column{3} = {c},
  column{5} = {c},
  column{7} = {c},
  cell{1}{2} = {c=2}{},
  cell{1}{4} = {c=2}{},
  cell{1}{6} = {c=2}{},
  cell{2}{2} = {c},
  cell{2}{4} = {c},
  cell{2}{6} = {c},
  cell{3}{2} = {c},
  cell{3}{4} = {c},
  cell{3}{6} = {c},
  cell{4}{2} = {c},
  cell{4}{4} = {c},
  cell{4}{6} = {c},
  cell{5}{2} = {c},
  cell{5}{4} = {c},
  cell{5}{6} = {c},
  cell{6}{2} = {c},
  cell{6}{4} = {c},
  cell{6}{6} = {c},
  cell{7}{2} = {c},
  cell{7}{4} = {c},
  cell{7}{6} = {c},
  cell{8}{2} = {c},
  cell{8}{4} = {c},
  cell{8}{6} = {c},
  hline{1,3,9} = {-}{},
  hline{2} = {2-7}{},
}
\textbf{Model} & \textbf{FashionVC} &                 & \textbf{ExpReduced} &                 & \textbf{FashionTaobaoTB} &                 \\
               & \textit{AUC}       & \textit{MRR}    & \textit{AUC}        & \textit{MRR}    & \textit{AUC}              & \textit{MRR}    \\
\textbf{\cgcrm} & \textbf{0.7684}    & \textbf{0.5197} & \textbf{0.6817}     & \textbf{0.4353} & \textbf{0.8500}           & \textbf{0.6463} \\
MGCM           & ~ 0.5667~~         & ~ 0.3455~~      & ~ 0.5048~~          & ~ 0.3173~~      & 0.7503                    & \uline{0.5255}  \\
Pix2PixCM      & 0.5611             & 0.3490          & 0.5177              & 0.2988          & 0.6981                    & 0.4668          \\
BPR-DAE        & \uline{0.6600}     & \uline{0.4236}  & \uline{0.6184}      & \uline{0.3660}  & \uline{0.7509}            & 0.5023          \\
Popularity            & 0.2074             & 0.2898          & 0.1957              & 0.2911          & 0.2416                    & 0.2931          \\
Random         & 0.5051             & 0.2883          & 0.5051              & 0.2883          & 0.4997                    & 0.2930          
\end{tblr}
\end{table}

%% file: figures/distribution.tex
\begin{figure*}[h]
\small
    
\begin{subfigure}[b]{0.3\textwidth}

\begin{tikzpicture}[scale=0.5]
\begin{axis}[
    axis lines=left,
    ybar,
    bar width=1,
    xlabel={Number of paired tops},    ylabel={Frequency},xlabel style={font=\Large}, 
    ylabel style={font=\Large}, 
    xtick={2, 4, 6, 8, 10, 12, 14, 16, 18, 20, 22, 24},
    xticklabel style={font=\Large}, %
    yticklabel style={font=\Large}, %
    enlarge x limits=0.05,
    ymax=9200
            ]
\addplot [fill=orange] table[y index=1]  {
1 8812
2 1434
3 417
4 210
5 91
6 92
7 60
8 44
9 26
10 29
11 17
12 9
13 10
14 9
15 3
16 1
17 2
18 3
19 0
20 3
21 2
22 2
23 0
24 0
                            };
\end{axis}
\label{fig:dist_fvc}
\end{tikzpicture}
\caption{FashionVC}
\end{subfigure}
\hfil
\begin{subfigure}[b]{0.3\textwidth}
\begin{tikzpicture}[scale=0.5]
\begin{axis}[
    axis lines=left,
    ybar,
    bar width=1,
    xlabel={Number of paired tops},    ylabel={Frequency},
    xlabel style={font=\Large}, 
    ylabel style={font=\Large}, 
    xtick={2, 4, 6, 8, 10, 12, 14, 16, 18, 20, 22, 24},
    xticklabel style={font=\Large}, %
    yticklabel style={font=\Large}, %
    enlarge x limits=0.05,
    ymax=5900
            ]
\addplot [fill=orange] table[y index=1]  {
1 5560
2 2195
3 987
4 428
5 220
6 75
7 47
8 27
9 10
10 10
11 4
12 3
13 1
14 0
15 2
16 2
17 2
18 0
19 1
20 0
21 0
22 0
23 0
24 1
                            };

\end{axis}
\label{fig:dist_expred}
\end{tikzpicture}
\caption{ExpReduced}
\end{subfigure}
\hfil
\pgfplotsset{y tick label style={scaled y ticks = false,
font=\Large,
/pgf/number format/.cd, fixed, fixed zerofill, int detect,precision=3}}
\begin{subfigure}[b]{0.3\textwidth}

\begin{tikzpicture}[scale=0.5]
\begin{axis}[
    axis lines=left,
    ybar,
    bar width=1,
    xlabel={Number of paired tops},    ylabel={Frequency},
    xlabel style={font=\Large}, 
    ylabel style={font=\Large}, 
    xtick={2, 4, 6, 8, 10, 12, 14, 16, 18, 20, 22, 24},
    xticklabel style={font=\Large}, %
    ytick={2000,6000,10000,14000},
    enlarge x limits=0.05,
    ymax=14900
            ]
\addplot [fill=orange] table[y index=1]  {
1 14484
2 4589
3 2034
4 1141
5 746
6 483
7 408
8 288
9 206
10 138
11 134
12 126
13 77
14 92
15 76
16 58
17 63
18 40
19 49
20 34
21 23
22 32
23 29
24 23
                            };

\end{axis}
\label{fig:dist_ftb}
\end{tikzpicture}
\caption{FashionTaobaoTB}
\end{subfigure}
\caption{Distribution of pairs across all datasets.}
\label{fig:pop_dist}
\end{figure*} 

%% file: tables/ablation.tex
\begin{table}[h!]
\centering
\caption{Ablation Study results. For each metric and dataset, \textbf{boldface} and \uline{underlined} indicate best and second-to-best values.}
\label{table:ablation_study}
\begin{tblr}{
  column{even} = {c},
  column{3} = {c},
  column{5} = {c},
  column{7} = {c},
  cell{1}{2} = {c=2}{},
  cell{1}{4} = {c=2}{},
  cell{1}{6} = {c=2}{},
  hline{1,3,6} = {-}{},
  hline{2} = {2-7}{},
}
\textbf{\textbf{\textbf{\textbf{Model}}}} & \textbf{~ FashionVC ~}   &                  & \textbf{~ ~ExpReduced ~~} &                 & \textbf{FashionTaobaoTB } &                          \\
                                          & \textit{AUC}             & \textit{MRR}     & \textit{AUC}              & \textit{MRR}    & \textit{AUC}              & \textit{MRR}             \\
\cgcrm-\textit{w/o BPR}                    & 0.7502                   & ~\textbf{0.0335} & 0.6846                    & \textbf{0.0193} & \uline{0.8382}            & \textbf{\textbf{0.0134}} \\
\cgcrm-\textit{w/o \nce}                       & \uline{0.7285}           & 0.0286           & \uline{0.6498}            & \uline{0.0178}  & 0.7888                    & 0.0060                   \\
\cgcrm                                     & \textbf{\textbf{0.7621}} & \uline{0.0307}   & \textbf{\textbf{0.7295}}  & 0.0151          & \textbf{\textbf{0.8491}}  & \uline{0.0126}           
\end{tblr}
\end{table}

%% file: figures/alpha.tex
\begin{figure}[!t]
\captionsetup[subfigure]{font=small,labelfont=small}
\small
\centering
    \subfloat[Effects of changing the value of $\alpha$.]{

\begin{tikzpicture}
\begin{scope}[scale=0.57, transform shape]
\begin{axis}[
    legend columns=3,
    legend style={at={(0.5,-0.25)},anchor=north},
    yticklabel style={
        /pgf/number format/.cd,
        fixed, fixed zerofill,
        precision=2
    },
    scaled y ticks=false,
     xticklabels={0.25,0.5,0.75,1},
     xtick={0.25,0.5,0.75,1},
     xtick pos=left,
     ytick pos=left,
     ymax=0.86,
     ymin=0.80,
     yticklabels={0.81,0.82,0.83,0.84,0.85},
     ytick={0.81,0.82,0.83,0.84,0.85},
     ylabel={AUC}
    ]
\addplot +[blue,line width=0.22mm,mark=o,mark size=3pt] table [x=alpha, y=AUC, col sep=comma] {figures/alpha.tsv};

\end{axis}
\end{scope}
\label{fig:hp_alpha}
\end{tikzpicture}

\quad

\begin{tikzpicture}
\begin{scope}[scale=0.57, transform shape]
\begin{axis}[
    legend columns=3,
    legend style={at={(0.5,-0.25)},anchor=north},
    yticklabel style={
        /pgf/number format/.cd,
        fixed, fixed zerofill,
        precision=2
    },
    scaled y ticks=false,
     xticklabels={0.25,0.5,0.75,1},
     xtick={0.25,0.5,0.75,1},
     xtick pos=left,
     ytick pos=left,
     ymax=0.017,
     ymin=0.007,
     yticklabels={0.008,0.010,0.012,0.014,0.016},
     ytick={0.008,0.010,0.012,0.014,0.016},
     ylabel={MRR}
    ]
\addplot +[orange,line width=0.22mm,mark=+,mark size=3pt] table [x=alpha, y=MRR, col sep=comma] {figures/alpha.tsv};

\end{axis}
\end{scope}
\end{tikzpicture}
}
\vspace{0.3cm}

\subfloat[Effects of changing the value of $\beta$.]{

\begin{tikzpicture}
\begin{scope}[scale=0.57, transform shape]
\begin{axis}[
    legend columns=3,
    legend style={at={(0.5,-0.25)},anchor=north},
    yticklabel style={
        /pgf/number format/.cd,
        fixed, fixed zerofill,
        precision=2
    },
    scaled y ticks=false,
     xticklabels={0.25,0.5,0.75,1},
     xtick={0.25,0.5,0.75,1},
     xtick pos=left,
     ytick pos=left,
     ymax=0.86,
     ymin=0.80,
     yticklabels={0.81,0.82,0.83,0.84,0.85},
     ytick={0.81,0.82,0.83,0.84,0.85},
     ylabel={AUC}
    ]
\addplot +[blue,line width=0.22mm,mark=o,mark size=3pt] table [x=beta, y=AUC, col sep=comma] {figures/beta.tsv};

\end{axis}
\end{scope}
\label{fig:hp_beta}
\end{tikzpicture}

\quad

\begin{tikzpicture}
\begin{scope}[scale=0.57, transform shape]
\begin{axis}[
    legend columns=3,
    legend style={at={(0.5,-0.25)},anchor=north},
    yticklabel style={
        /pgf/number format/.cd,
        fixed, fixed zerofill,
        precision=2
    },
    scaled y ticks=false,
     xticklabels={0.25,0.5,0.75,1},
     xtick={0.25,0.5,0.75,1},
     xtick pos=left,
     ytick pos=left,
     ymax=0.017,
     ymin=0.007,
     yticklabels={0.008,0.010,0.012,0.014,0.016},
     ytick={0.008,0.010,0.012,0.014,0.016},
     ylabel={MRR}
    ]
\addplot +[orange,line width=0.22mm,mark=+,mark size=3pt] table [x=beta, y=MRR, col sep=comma] {figures/beta.tsv};

\end{axis}
\end{scope}
\end{tikzpicture}
}

\vspace{0.3cm}

\subfloat[Impact of varying the $\tau$ value.]{

\begin{tikzpicture}
\begin{scope}[scale=0.57, transform shape]
\begin{axis}[
    legend columns=3,
    legend style={at={(0.5,-0.25)},anchor=north},
    yticklabel style={
        /pgf/number format/.cd,
        fixed, fixed zerofill,
        precision=2
    },
    scaled y ticks=false,
     xticklabels={0.005,0.01,0.05,0.1,0.5},
     xtick={1,2,3,4,5},
     xtick pos=left,
     ytick pos=left,
     ymax=0.86,
     ymin=0.80,
     yticklabels={0.81,0.82,0.83,0.84,0.85},
     ytick={0.81,0.82,0.83,0.84,0.85},
     ylabel={AUC}
    ]
\addplot +[blue,line width=0.22mm,mark=o,mark size=3pt] table [x=tau, y=AUC, col sep=comma] {figures/tau.tsv};

\end{axis}
\end{scope}
\label{fig:hp_tau}
\end{tikzpicture}

\quad

\begin{tikzpicture}
\begin{scope}[scale=0.57, transform shape]
\begin{axis}[
    legend columns=3,
    legend style={at={(0.5,-0.25)},anchor=north},
    yticklabel style={
        /pgf/number format/.cd,
        fixed, fixed zerofill,
        precision=2
    },
    scaled y ticks=false,
     xticklabels={0.005,0.01,0.05,0.1,0.5},
     xtick={1,2,3,4,5},
     xtick pos=left,
     ytick pos=left,
     ymax=0.017,
     ymin=0.007,
     yticklabels={0.008,0.010,0.012,0.014,0.016},
     ytick={0.008,0.010,0.012,0.014,0.016},
     ylabel={MRR}
    ]
\addplot +[orange,line width=0.22mm,mark=+,mark size=3pt] table [x=tau, y=MRR, col sep=comma] {figures/tau.tsv};

\end{axis}
\end{scope}
\end{tikzpicture}
}
\caption{Scatter plots illustrating variations in terms of AUC and MRR in response to adjustments of the loss parameters on the FashionTaobaoTB dataset.}
\label{fig:hp_sensitivity}
\end{figure}

%% file: src/conclusions.tex
\section{Conclusions}\label{sec:6}
In this study, we have advanced the field of fashion compatibility modeling and fashion image retrieval by introducing a novel compatibility schema, \cgcrm. This schema integrates realistic generations into both the compatibility modeling and retrieval processes, thereby enhancing the overall effectiveness of these tasks. To the best of our knowledge, \cgcrm is the first to introduce a composed retrieval pipeline that leverages generated items as conditioning signals in a two-stage approach. This innovative method has demonstrated marked improvements in reliability and stability, particularly in scenarios characterized by limited training data.

Notably, \cgcrm not only pioneers a two-stage approach but also introduces a new framework that redefines the compatibility modeling and top-bottom retrieval tasks as a composed image retrieval problem. Unlike traditional methods that rely on an image-text query pair, our model is the first to utilize a composed query comprising two images (e.g., a top image and a template image). Furthermore, \cgcrm extends the concept of composed image retrieval across entirely different domains (e.g., top and bottom distributions), as opposed to previous works that focus on retrieving items with minor modifications based on a textual query. A remarkable outcome of our study is the ability of \cgcrm to generate realistic templates that effectively address key challenges in these tasks, a capability not observed in existing baselines. The experimental results also reveal a critical gap in the literature concerning applying generative models in contexts with limited data. 

Looking forward, several promising directions for future research emerge. One potential approach is the integration of textual signals within the developed framework to enhance retrieval performance further. Careful investigation will be needed to select the optimal point for incorporating these signals — whether at the GAN level or during the second stage of the \cgcrm. Additionally, exploring the integration of personalization features into the retrieval pipeline presents another promising area of research, to identify the most effective phase for their inclusion. These enhancements have the potential to further elevate the performance of fashion compatibility modeling and retrieval systems, providing users with more tailored and contextually relevant results.

%% file: main.bbl

\begin{thebibliography}{61}


\ifx \showCODEN    \undefined \def \showCODEN     #1{\unskip}     \fi
\ifx \showDOI      \undefined \def \showDOI       #1{#1}\fi
\ifx \showISBNx    \undefined \def \showISBNx     #1{\unskip}     \fi
\ifx \showISBNxiii \undefined \def \showISBNxiii  #1{\unskip}     \fi
\ifx \showISSN     \undefined \def \showISSN      #1{\unskip}     \fi
\ifx \showLCCN     \undefined \def \showLCCN      #1{\unskip}     \fi
\ifx \shownote     \undefined \def \shownote      #1{#1}          \fi
\ifx \showarticletitle \undefined \def \showarticletitle #1{#1}   \fi
\ifx \showURL      \undefined \def \showURL       {\relax}        \fi
\providecommand\bibfield[2]{#2}
\providecommand\bibinfo[2]{#2}
\providecommand\natexlab[1]{#1}
\providecommand\showeprint[2][]{arXiv:#2}

\bibitem[Arjovsky and Bottou(2017)]%
        {DBLP:conf/iclr/ArjovskyB17}
\bibfield{author}{\bibinfo{person}{Mart{\'{\i}}n Arjovsky} {and} \bibinfo{person}{L{\'{e}}on Bottou}.} \bibinfo{year}{2017}\natexlab{}.
\newblock \showarticletitle{Towards Principled Methods for Training Generative Adversarial Networks}. In \bibinfo{booktitle}{\emph{{ICLR}}}. \bibinfo{publisher}{OpenReview.net}.
\newblock


\bibitem[Arjovsky et~al\mbox{.}(2017)]%
        {DBLP:conf/icml/ArjovskyCB17}
\bibfield{author}{\bibinfo{person}{Mart{\'{\i}}n Arjovsky}, \bibinfo{person}{Soumith Chintala}, {and} \bibinfo{person}{L{\'{e}}on Bottou}.} \bibinfo{year}{2017}\natexlab{}.
\newblock \showarticletitle{Wasserstein Generative Adversarial Networks}. In \bibinfo{booktitle}{\emph{{ICML}}} \emph{(\bibinfo{series}{Proceedings of Machine Learning Research}, Vol.~\bibinfo{volume}{70})}. \bibinfo{publisher}{{PMLR}}, \bibinfo{pages}{214--223}.
\newblock


\bibitem[Baldrati et~al\mbox{.}(2023)]%
        {DBLP:conf/iccv/BaldratiA0B23}
\bibfield{author}{\bibinfo{person}{Alberto Baldrati}, \bibinfo{person}{Lorenzo Agnolucci}, \bibinfo{person}{Marco Bertini}, {and} \bibinfo{person}{Alberto~Del Bimbo}.} \bibinfo{year}{2023}\natexlab{}.
\newblock \showarticletitle{Zero-Shot Composed Image Retrieval with Textual Inversion}. In \bibinfo{booktitle}{\emph{{ICCV}}}. \bibinfo{publisher}{{IEEE}}, \bibinfo{pages}{15292--15301}.
\newblock


\bibitem[Baldrati et~al\mbox{.}(2022)]%
        {DBLP:conf/cvpr/BaldratiBUB22a}
\bibfield{author}{\bibinfo{person}{Alberto Baldrati}, \bibinfo{person}{Marco Bertini}, \bibinfo{person}{Tiberio Uricchio}, {and} \bibinfo{person}{Alberto~Del Bimbo}.} \bibinfo{year}{2022}\natexlab{}.
\newblock \showarticletitle{Effective conditioned and composed image retrieval combining CLIP-based features}. In \bibinfo{booktitle}{\emph{{CVPR}}}. \bibinfo{publisher}{{IEEE}}, \bibinfo{pages}{21434--21442}.
\newblock


\bibitem[Baldrati et~al\mbox{.}(2024)]%
        {DBLP:journals/tomccap/BaldratiBUB24}
\bibfield{author}{\bibinfo{person}{Alberto Baldrati}, \bibinfo{person}{Marco Bertini}, \bibinfo{person}{Tiberio Uricchio}, {and} \bibinfo{person}{Alberto~Del Bimbo}.} \bibinfo{year}{2024}\natexlab{}.
\newblock \showarticletitle{Composed Image Retrieval using Contrastive Learning and Task-oriented CLIP-based Features}.
\newblock \bibinfo{journal}{\emph{{ACM} Trans. Multim. Comput. Commun. Appl.}} \bibinfo{volume}{20}, \bibinfo{number}{3} (\bibinfo{year}{2024}), \bibinfo{pages}{62:1--62:24}.
\newblock


\bibitem[Berthelot et~al\mbox{.}(2024)]%
        {berthelot2024estimating}
\bibfield{author}{\bibinfo{person}{Adrien Berthelot}, \bibinfo{person}{Eddy Caron}, \bibinfo{person}{Mathilde Jay}, {and} \bibinfo{person}{Laurent Lef{\`e}vre}.} \bibinfo{year}{2024}\natexlab{}.
\newblock \showarticletitle{Estimating the environmental impact of Generative-AI services using an LCA-based methodology}.
\newblock \bibinfo{journal}{\emph{Procedia CIRP}}  \bibinfo{volume}{122} (\bibinfo{year}{2024}), \bibinfo{pages}{707--712}.
\newblock


\bibitem[Bibas et~al\mbox{.}(2023)]%
        {DBLP:conf/www/BibasSJ23}
\bibfield{author}{\bibinfo{person}{Koby Bibas}, \bibinfo{person}{Oren~Sar Shalom}, {and} \bibinfo{person}{Dietmar Jannach}.} \bibinfo{year}{2023}\natexlab{}.
\newblock \showarticletitle{Semi-supervised Adversarial Learning for Complementary Item Recommendation}. In \bibinfo{booktitle}{\emph{{WWW}}}. \bibinfo{publisher}{{ACM}}, \bibinfo{pages}{1804--1812}.
\newblock


\bibitem[Chen et~al\mbox{.}(2017)]%
        {DBLP:conf/sigir/ChenZ0NLC17}
\bibfield{author}{\bibinfo{person}{Jingyuan Chen}, \bibinfo{person}{Hanwang Zhang}, \bibinfo{person}{Xiangnan He}, \bibinfo{person}{Liqiang Nie}, \bibinfo{person}{Wei Liu}, {and} \bibinfo{person}{Tat{-}Seng Chua}.} \bibinfo{year}{2017}\natexlab{}.
\newblock \showarticletitle{Attentive Collaborative Filtering: Multimedia Recommendation with Item- and Component-Level Attention}. In \bibinfo{booktitle}{\emph{{SIGIR}}}. \bibinfo{publisher}{{ACM}}, \bibinfo{pages}{335--344}.
\newblock


\bibitem[Chen et~al\mbox{.}(2019)]%
        {DBLP:conf/kdd/ChenHXGGSLPZZ19}
\bibfield{author}{\bibinfo{person}{Wen Chen}, \bibinfo{person}{Pipei Huang}, \bibinfo{person}{Jiaming Xu}, \bibinfo{person}{Xin Guo}, \bibinfo{person}{Cheng Guo}, \bibinfo{person}{Fei Sun}, \bibinfo{person}{Chao Li}, \bibinfo{person}{Andreas Pfadler}, \bibinfo{person}{Huan Zhao}, {and} \bibinfo{person}{Binqiang Zhao}.} \bibinfo{year}{2019}\natexlab{}.
\newblock \showarticletitle{{POG:} Personalized Outfit Generation for Fashion Recommendation at Alibaba iFashion}. In \bibinfo{booktitle}{\emph{{KDD}}}. \bibinfo{publisher}{{ACM}}, \bibinfo{pages}{2662--2670}.
\newblock


\bibitem[Cui et~al\mbox{.}(2019)]%
        {DBLP:conf/www/CuiLWZW19}
\bibfield{author}{\bibinfo{person}{Zeyu Cui}, \bibinfo{person}{Zekun Li}, \bibinfo{person}{Shu Wu}, \bibinfo{person}{Xiaoyu Zhang}, {and} \bibinfo{person}{Liang Wang}.} \bibinfo{year}{2019}\natexlab{}.
\newblock \showarticletitle{Dressing as a Whole: Outfit Compatibility Learning Based on Node-wise Graph Neural Networks}. In \bibinfo{booktitle}{\emph{{WWW}}}. \bibinfo{publisher}{{ACM}}, \bibinfo{pages}{307--317}.
\newblock


\bibitem[Deldjoo et~al\mbox{.}(2024)]%
        {DBLP:journals/csur/DeldjooNRMPBN24}
\bibfield{author}{\bibinfo{person}{Yashar Deldjoo}, \bibinfo{person}{Fatemeh Nazary}, \bibinfo{person}{Arnau Ramisa}, \bibinfo{person}{Julian~J. McAuley}, \bibinfo{person}{Giovanni Pellegrini}, \bibinfo{person}{Alejandro Bellog{\'{\i}}n}, {and} \bibinfo{person}{Tommaso~Di Noia}.} \bibinfo{year}{2024}\natexlab{}.
\newblock \showarticletitle{A Review of Modern Fashion Recommender Systems}.
\newblock \bibinfo{journal}{\emph{{ACM} Comput. Surv.}} \bibinfo{volume}{56}, \bibinfo{number}{4} (\bibinfo{year}{2024}), \bibinfo{pages}{87:1--87:37}.
\newblock


\bibitem[Deldjoo et~al\mbox{.}(2021)]%
        {DBLP:conf/cvpr/DeldjooNMM21}
\bibfield{author}{\bibinfo{person}{Yashar Deldjoo}, \bibinfo{person}{Tommaso~Di Noia}, \bibinfo{person}{Daniele Malitesta}, {and} \bibinfo{person}{Felice~Antonio Merra}.} \bibinfo{year}{2021}\natexlab{}.
\newblock \showarticletitle{A Study on the Relative Importance of Convolutional Neural Networks in Visually-Aware Recommender Systems}. In \bibinfo{booktitle}{\emph{{CVPR} Workshops}}. \bibinfo{publisher}{Computer Vision Foundation / {IEEE}}, \bibinfo{pages}{3961--3967}.
\newblock


\bibitem[Dhariwal and Nichol(2021)]%
        {DBLP:conf/nips/DhariwalN21}
\bibfield{author}{\bibinfo{person}{Prafulla Dhariwal} {and} \bibinfo{person}{Alexander~Quinn Nichol}.} \bibinfo{year}{2021}\natexlab{}.
\newblock \showarticletitle{Diffusion Models Beat GANs on Image Synthesis}. In \bibinfo{booktitle}{\emph{NeurIPS}}. \bibinfo{pages}{8780--8794}.
\newblock


\bibitem[El{-}Kaddoury et~al\mbox{.}(2019)]%
        {DBLP:conf/mspn/El-KaddouryMH19}
\bibfield{author}{\bibinfo{person}{Mohamed El{-}Kaddoury}, \bibinfo{person}{Abdelhak Mahmoudi}, {and} \bibinfo{person}{Mohamed~Majid Himmi}.} \bibinfo{year}{2019}\natexlab{}.
\newblock \showarticletitle{Deep Generative Models for Image Generation: {A} Practical Comparison Between Variational Autoencoders and Generative Adversarial Networks}. In \bibinfo{booktitle}{\emph{{MSPN}}}. \bibinfo{publisher}{Springer}.
\newblock


\bibitem[Fawcett(2006)]%
        {DBLP:journals/prl/Fawcett06}
\bibfield{author}{\bibinfo{person}{Tom Fawcett}.} \bibinfo{year}{2006}\natexlab{}.
\newblock \showarticletitle{An introduction to {ROC} analysis}.
\newblock \bibinfo{journal}{\emph{Pattern Recognit. Lett.}} \bibinfo{volume}{27}, \bibinfo{number}{8} (\bibinfo{year}{2006}), \bibinfo{pages}{861--874}.
\newblock


\bibitem[Feng et~al\mbox{.}(2023)]%
        {DBLP:journals/corr/abs-2312-12273}
\bibfield{author}{\bibinfo{person}{Chun{-}Mei Feng}, \bibinfo{person}{Yang Bai}, \bibinfo{person}{Tao Luo}, \bibinfo{person}{Zhen Li}, \bibinfo{person}{Salman Khan}, \bibinfo{person}{Wangmeng Zuo}, \bibinfo{person}{Xinxing Xu}, \bibinfo{person}{Rick Siow~Mong Goh}, {and} \bibinfo{person}{Yong Liu}.} \bibinfo{year}{2023}\natexlab{}.
\newblock \showarticletitle{{VQA4CIR:} Boosting Composed Image Retrieval with Visual Question Answering}.
\newblock \bibinfo{journal}{\emph{CoRR}}  \bibinfo{volume}{abs/2312.12273} (\bibinfo{year}{2023}).
\newblock


\bibitem[Feng et~al\mbox{.}(2024)]%
        {feng2024improving}
\bibfield{author}{\bibinfo{person}{Zhangchi Feng}, \bibinfo{person}{Richong Zhang}, {and} \bibinfo{person}{Zhijie Nie}.} \bibinfo{year}{2024}\natexlab{}.
\newblock \showarticletitle{Improving Composed Image Retrieval via Contrastive Learning with Scaling Positives and Negatives}.
\newblock \bibinfo{journal}{\emph{arXiv preprint arXiv:2404.11317}} (\bibinfo{year}{2024}).
\newblock


\bibitem[Goodfellow(2017)]%
        {DBLP:journals/corr/Goodfellow17}
\bibfield{author}{\bibinfo{person}{Ian~J. Goodfellow}.} \bibinfo{year}{2017}\natexlab{}.
\newblock \showarticletitle{{NIPS} 2016 Tutorial: Generative Adversarial Networks}.
\newblock \bibinfo{journal}{\emph{CoRR}}  \bibinfo{volume}{abs/1701.00160} (\bibinfo{year}{2017}).
\newblock


\bibitem[Goodfellow et~al\mbox{.}(2014)]%
        {DBLP:journals/corr/GoodfellowPMXWOCB14}
\bibfield{author}{\bibinfo{person}{Ian~J. Goodfellow}, \bibinfo{person}{Jean Pouget{-}Abadie}, \bibinfo{person}{Mehdi Mirza}, \bibinfo{person}{Bing Xu}, \bibinfo{person}{David Warde{-}Farley}, \bibinfo{person}{Sherjil Ozair}, \bibinfo{person}{Aaron~C. Courville}, {and} \bibinfo{person}{Yoshua Bengio}.} \bibinfo{year}{2014}\natexlab{}.
\newblock \showarticletitle{Generative Adversarial Networks}.
\newblock \bibinfo{journal}{\emph{CoRR}}  \bibinfo{volume}{abs/1406.2661} (\bibinfo{year}{2014}).
\newblock


\bibitem[Gulrajani et~al\mbox{.}(2017)]%
        {DBLP:conf/nips/GulrajaniAADC17}
\bibfield{author}{\bibinfo{person}{Ishaan Gulrajani}, \bibinfo{person}{Faruk Ahmed}, \bibinfo{person}{Mart{\'{\i}}n Arjovsky}, \bibinfo{person}{Vincent Dumoulin}, {and} \bibinfo{person}{Aaron~C. Courville}.} \bibinfo{year}{2017}\natexlab{}.
\newblock \showarticletitle{Improved Training of Wasserstein GANs}. In \bibinfo{booktitle}{\emph{{NIPS}}}. \bibinfo{pages}{5767--5777}.
\newblock


\bibitem[Hao et~al\mbox{.}(2020)]%
        {DBLP:conf/cikm/HaoZLDFSW20}
\bibfield{author}{\bibinfo{person}{Junheng Hao}, \bibinfo{person}{Tong Zhao}, \bibinfo{person}{Jin Li}, \bibinfo{person}{Xin~Luna Dong}, \bibinfo{person}{Christos Faloutsos}, \bibinfo{person}{Yizhou Sun}, {and} \bibinfo{person}{Wei Wang}.} \bibinfo{year}{2020}\natexlab{}.
\newblock \showarticletitle{P-Companion: {A} Principled Framework for Diversified Complementary Product Recommendation}. In \bibinfo{booktitle}{\emph{{CIKM}}}. \bibinfo{publisher}{{ACM}}, \bibinfo{pages}{2517--2524}.
\newblock


\bibitem[He et~al\mbox{.}(2016)]%
        {DBLP:conf/cvpr/HeZRS16}
\bibfield{author}{\bibinfo{person}{Kaiming He}, \bibinfo{person}{Xiangyu Zhang}, \bibinfo{person}{Shaoqing Ren}, {and} \bibinfo{person}{Jian Sun}.} \bibinfo{year}{2016}\natexlab{}.
\newblock \showarticletitle{Deep Residual Learning for Image Recognition}. In \bibinfo{booktitle}{\emph{{CVPR}}}. \bibinfo{publisher}{{IEEE} Computer Society}, \bibinfo{pages}{770--778}.
\newblock


\bibitem[He and McAuley(2016)]%
        {DBLP:conf/aaai/HeM16}
\bibfield{author}{\bibinfo{person}{Ruining He} {and} \bibinfo{person}{Julian~J. McAuley}.} \bibinfo{year}{2016}\natexlab{}.
\newblock \showarticletitle{{VBPR:} Visual Bayesian Personalized Ranking from Implicit Feedback}. In \bibinfo{booktitle}{\emph{{AAAI}}}. \bibinfo{publisher}{{AAAI} Press}, \bibinfo{pages}{144--150}.
\newblock


\bibitem[Ho et~al\mbox{.}(2020)]%
        {DBLP:conf/nips/HoJA20}
\bibfield{author}{\bibinfo{person}{Jonathan Ho}, \bibinfo{person}{Ajay Jain}, {and} \bibinfo{person}{Pieter Abbeel}.} \bibinfo{year}{2020}\natexlab{}.
\newblock \showarticletitle{Denoising Diffusion Probabilistic Models}. In \bibinfo{booktitle}{\emph{NeurIPS}}.
\newblock


\bibitem[Hoffmann et~al\mbox{.}(2022)]%
        {DBLP:conf/aaai/HoffmannBGBN22}
\bibfield{author}{\bibinfo{person}{David~T. Hoffmann}, \bibinfo{person}{Nadine Behrmann}, \bibinfo{person}{Juergen Gall}, \bibinfo{person}{Thomas Brox}, {and} \bibinfo{person}{Mehdi Noroozi}.} \bibinfo{year}{2022}\natexlab{}.
\newblock \showarticletitle{Ranking Info Noise Contrastive Estimation: Boosting Contrastive Learning via Ranked Positives}. In \bibinfo{booktitle}{\emph{{AAAI}}}. \bibinfo{publisher}{{AAAI} Press}, \bibinfo{pages}{897--905}.
\newblock


\bibitem[Isola et~al\mbox{.}(2017)]%
        {DBLP:conf/cvpr/IsolaZZE17}
\bibfield{author}{\bibinfo{person}{Phillip Isola}, \bibinfo{person}{Jun{-}Yan Zhu}, \bibinfo{person}{Tinghui Zhou}, {and} \bibinfo{person}{Alexei~A. Efros}.} \bibinfo{year}{2017}\natexlab{}.
\newblock \showarticletitle{Image-to-Image Translation with Conditional Adversarial Networks}. In \bibinfo{booktitle}{\emph{{CVPR}}}. \bibinfo{publisher}{{IEEE} Computer Society}, \bibinfo{pages}{5967--5976}.
\newblock


\bibitem[Kang et~al\mbox{.}(2017)]%
        {DBLP:conf/icdm/KangFWM17}
\bibfield{author}{\bibinfo{person}{Wang{-}Cheng Kang}, \bibinfo{person}{Chen Fang}, \bibinfo{person}{Zhaowen Wang}, {and} \bibinfo{person}{Julian~J. McAuley}.} \bibinfo{year}{2017}\natexlab{}.
\newblock \showarticletitle{Visually-Aware Fashion Recommendation and Design with Generative Image Models}. In \bibinfo{booktitle}{\emph{{ICDM}}}. \bibinfo{publisher}{{IEEE} Computer Society}, \bibinfo{pages}{207--216}.
\newblock


\bibitem[Karras et~al\mbox{.}(2022)]%
        {DBLP:conf/nips/KarrasAAL22}
\bibfield{author}{\bibinfo{person}{Tero Karras}, \bibinfo{person}{Miika Aittala}, \bibinfo{person}{Timo Aila}, {and} \bibinfo{person}{Samuli Laine}.} \bibinfo{year}{2022}\natexlab{}.
\newblock \showarticletitle{Elucidating the Design Space of Diffusion-Based Generative Models}. In \bibinfo{booktitle}{\emph{NeurIPS}}.
\newblock


\bibitem[Kingma and Welling(2014)]%
        {DBLP:journals/corr/KingmaW13}
\bibfield{author}{\bibinfo{person}{Diederik~P. Kingma} {and} \bibinfo{person}{Max Welling}.} \bibinfo{year}{2014}\natexlab{}.
\newblock \showarticletitle{Auto-Encoding Variational Bayes}. In \bibinfo{booktitle}{\emph{{ICLR}}}.
\newblock


\bibitem[Koren et~al\mbox{.}(2009)]%
        {DBLP:journals/computer/KorenBV09}
\bibfield{author}{\bibinfo{person}{Yehuda Koren}, \bibinfo{person}{Robert~M. Bell}, {and} \bibinfo{person}{Chris Volinsky}.} \bibinfo{year}{2009}\natexlab{}.
\newblock \showarticletitle{Matrix Factorization Techniques for Recommender Systems}.
\newblock \bibinfo{journal}{\emph{Computer}} \bibinfo{volume}{42}, \bibinfo{number}{8} (\bibinfo{year}{2009}), \bibinfo{pages}{30--37}.
\newblock


\bibitem[Ledig et~al\mbox{.}(2017)]%
        {DBLP:conf/cvpr/LedigTHCCAATTWS17}
\bibfield{author}{\bibinfo{person}{Christian Ledig}, \bibinfo{person}{Lucas Theis}, \bibinfo{person}{Ferenc Huszar}, \bibinfo{person}{Jose Caballero}, \bibinfo{person}{Andrew Cunningham}, \bibinfo{person}{Alejandro Acosta}, \bibinfo{person}{Andrew~P. Aitken}, \bibinfo{person}{Alykhan Tejani}, \bibinfo{person}{Johannes Totz}, \bibinfo{person}{Zehan Wang}, {and} \bibinfo{person}{Wenzhe Shi}.} \bibinfo{year}{2017}\natexlab{}.
\newblock \showarticletitle{Photo-Realistic Single Image Super-Resolution Using a Generative Adversarial Network}. In \bibinfo{booktitle}{\emph{{CVPR}}}. \bibinfo{publisher}{{IEEE} Computer Society}, \bibinfo{pages}{105--114}.
\newblock


\bibitem[Li et~al\mbox{.}(2020)]%
        {DBLP:conf/sigir/LiW0CXC20}
\bibfield{author}{\bibinfo{person}{Xingchen Li}, \bibinfo{person}{Xiang Wang}, \bibinfo{person}{Xiangnan He}, \bibinfo{person}{Long Chen}, \bibinfo{person}{Jun Xiao}, {and} \bibinfo{person}{Tat{-}Seng Chua}.} \bibinfo{year}{2020}\natexlab{}.
\newblock \showarticletitle{Hierarchical Fashion Graph Network for Personalized Outfit Recommendation}. In \bibinfo{booktitle}{\emph{{SIGIR}}}. \bibinfo{publisher}{{ACM}}, \bibinfo{pages}{159--168}.
\newblock


\bibitem[Lin et~al\mbox{.}(2019)]%
        {DBLP:conf/www/LinRCRMR19}
\bibfield{author}{\bibinfo{person}{Yujie Lin}, \bibinfo{person}{Pengjie Ren}, \bibinfo{person}{Zhumin Chen}, \bibinfo{person}{Zhaochun Ren}, \bibinfo{person}{Jun Ma}, {and} \bibinfo{person}{Maarten de Rijke}.} \bibinfo{year}{2019}\natexlab{}.
\newblock \showarticletitle{Improving Outfit Recommendation with Co-supervision of Fashion Generation}. In \bibinfo{booktitle}{\emph{{WWW}}}. \bibinfo{publisher}{{ACM}}, \bibinfo{pages}{1095--1105}.
\newblock


\bibitem[Lin et~al\mbox{.}(2020)]%
        {DBLP:journals/tkde/LinRCRMR20}
\bibfield{author}{\bibinfo{person}{Yujie Lin}, \bibinfo{person}{Pengjie Ren}, \bibinfo{person}{Zhumin Chen}, \bibinfo{person}{Zhaochun Ren}, \bibinfo{person}{Jun Ma}, {and} \bibinfo{person}{Maarten de Rijke}.} \bibinfo{year}{2020}\natexlab{}.
\newblock \showarticletitle{Explainable Outfit Recommendation with Joint Outfit Matching and Comment Generation}.
\newblock \bibinfo{journal}{\emph{{IEEE} TKDE}} \bibinfo{volume}{32}, \bibinfo{number}{8} (\bibinfo{year}{2020}), \bibinfo{pages}{1502--1516}.
\newblock


\bibitem[Liu et~al\mbox{.}(2020a)]%
        {DBLP:journals/ijon/LiuSCM20}
\bibfield{author}{\bibinfo{person}{Jinhuan Liu}, \bibinfo{person}{Xuemeng Song}, \bibinfo{person}{Zhumin Chen}, {and} \bibinfo{person}{Jun Ma}.} \bibinfo{year}{2020}\natexlab{a}.
\newblock \showarticletitle{{MGCM:} Multi-modal generative compatibility modeling for clothing matching}.
\newblock \bibinfo{journal}{\emph{Neurocomputing}}  \bibinfo{volume}{414} (\bibinfo{year}{2020}), \bibinfo{pages}{215--224}.
\newblock


\bibitem[Liu et~al\mbox{.}(2020b)]%
        {DBLP:conf/ijcai/LiuSRNT020}
\bibfield{author}{\bibinfo{person}{Jinhuan Liu}, \bibinfo{person}{Xuemeng Song}, \bibinfo{person}{Zhaochun Ren}, \bibinfo{person}{Liqiang Nie}, \bibinfo{person}{Zhaopeng Tu}, {and} \bibinfo{person}{Jun Ma}.} \bibinfo{year}{2020}\natexlab{b}.
\newblock \showarticletitle{Auxiliary Template-Enhanced Generative Compatibility Modeling}. In \bibinfo{booktitle}{\emph{{IJCAI}}}. \bibinfo{publisher}{ijcai.org}, \bibinfo{pages}{3508--3514}.
\newblock


\bibitem[Liu et~al\mbox{.}(2022)]%
        {DBLP:conf/iclr/Liu0LZ22}
\bibfield{author}{\bibinfo{person}{Luping Liu}, \bibinfo{person}{Yi Ren}, \bibinfo{person}{Zhijie Lin}, {and} \bibinfo{person}{Zhou Zhao}.} \bibinfo{year}{2022}\natexlab{}.
\newblock \showarticletitle{Pseudo Numerical Methods for Diffusion Models on Manifolds}. In \bibinfo{booktitle}{\emph{{ICLR}}}. \bibinfo{publisher}{OpenReview.net}.
\newblock


\bibitem[Liu et~al\mbox{.}(2017)]%
        {DBLP:conf/sigir/LiuWW17}
\bibfield{author}{\bibinfo{person}{Qiang Liu}, \bibinfo{person}{Shu Wu}, {and} \bibinfo{person}{Liang Wang}.} \bibinfo{year}{2017}\natexlab{}.
\newblock \showarticletitle{DeepStyle: Learning User Preferences for Visual Recommendation}. In \bibinfo{booktitle}{\emph{{SIGIR}}}. \bibinfo{publisher}{{ACM}}, \bibinfo{pages}{841--844}.
\newblock


\bibitem[Liu et~al\mbox{.}(2021)]%
        {DBLP:conf/iccv/0002OTG21}
\bibfield{author}{\bibinfo{person}{Zheyuan Liu}, \bibinfo{person}{Cristian~Rodriguez Opazo}, \bibinfo{person}{Damien Teney}, {and} \bibinfo{person}{Stephen Gould}.} \bibinfo{year}{2021}\natexlab{}.
\newblock \showarticletitle{Image Retrieval on Real-life Images with Pre-trained Vision-and-Language Models}. In \bibinfo{booktitle}{\emph{{ICCV}}}. \bibinfo{publisher}{{IEEE}}, \bibinfo{pages}{2105--2114}.
\newblock


\bibitem[Mescheder et~al\mbox{.}(2018)]%
        {DBLP:conf/icml/MeschederGN18}
\bibfield{author}{\bibinfo{person}{Lars~M. Mescheder}, \bibinfo{person}{Andreas Geiger}, {and} \bibinfo{person}{Sebastian Nowozin}.} \bibinfo{year}{2018}\natexlab{}.
\newblock \showarticletitle{Which Training Methods for GANs do actually Converge?}. In \bibinfo{booktitle}{\emph{{ICML}}} \emph{(\bibinfo{series}{Proceedings of Machine Learning Research}, Vol.~\bibinfo{volume}{80})}. \bibinfo{publisher}{{PMLR}}, \bibinfo{pages}{3478--3487}.
\newblock


\bibitem[Mirza and Osindero(2014)]%
        {DBLP:journals/corr/MirzaO14}
\bibfield{author}{\bibinfo{person}{Mehdi Mirza} {and} \bibinfo{person}{Simon Osindero}.} \bibinfo{year}{2014}\natexlab{}.
\newblock \showarticletitle{Conditional Generative Adversarial Nets}.
\newblock \bibinfo{journal}{\emph{CoRR}}  \bibinfo{volume}{abs/1411.1784} (\bibinfo{year}{2014}).
\newblock


\bibitem[Miyato et~al\mbox{.}(2018)]%
        {DBLP:conf/iclr/MiyatoKKY18}
\bibfield{author}{\bibinfo{person}{Takeru Miyato}, \bibinfo{person}{Toshiki Kataoka}, \bibinfo{person}{Masanori Koyama}, {and} \bibinfo{person}{Yuichi Yoshida}.} \bibinfo{year}{2018}\natexlab{}.
\newblock \showarticletitle{Spectral Normalization for Generative Adversarial Networks}. In \bibinfo{booktitle}{\emph{{ICLR}}}. \bibinfo{publisher}{OpenReview.net}.
\newblock


\bibitem[Park et~al\mbox{.}(2019)]%
        {DBLP:conf/cvpr/Park0WZ19}
\bibfield{author}{\bibinfo{person}{Taesung Park}, \bibinfo{person}{Ming{-}Yu Liu}, \bibinfo{person}{Ting{-}Chun Wang}, {and} \bibinfo{person}{Jun{-}Yan Zhu}.} \bibinfo{year}{2019}\natexlab{}.
\newblock \showarticletitle{Semantic Image Synthesis With Spatially-Adaptive Normalization}. In \bibinfo{booktitle}{\emph{{CVPR}}}. \bibinfo{publisher}{Computer Vision Foundation / {IEEE}}, \bibinfo{pages}{2337--2346}.
\newblock


\bibitem[Pascanu et~al\mbox{.}(2013)]%
        {DBLP:conf/icml/PascanuMB13}
\bibfield{author}{\bibinfo{person}{Razvan Pascanu}, \bibinfo{person}{Tom{\'{a}}s Mikolov}, {and} \bibinfo{person}{Yoshua Bengio}.} \bibinfo{year}{2013}\natexlab{}.
\newblock \showarticletitle{On the difficulty of training recurrent neural networks}. In \bibinfo{booktitle}{\emph{{ICML} {(3)}}} \emph{(\bibinfo{series}{{JMLR} Workshop and Conference Proceedings}, Vol.~\bibinfo{volume}{28})}. \bibinfo{publisher}{JMLR.org}, \bibinfo{pages}{1310--1318}.
\newblock


\bibitem[Rendle et~al\mbox{.}(2009)]%
        {DBLP:conf/uai/RendleFGS09}
\bibfield{author}{\bibinfo{person}{Steffen Rendle}, \bibinfo{person}{Christoph Freudenthaler}, \bibinfo{person}{Zeno Gantner}, {and} \bibinfo{person}{Lars Schmidt{-}Thieme}.} \bibinfo{year}{2009}\natexlab{}.
\newblock \showarticletitle{{BPR:} Bayesian Personalized Ranking from Implicit Feedback}. In \bibinfo{booktitle}{\emph{{UAI}}}. \bibinfo{publisher}{{AUAI} Press}, \bibinfo{pages}{452--461}.
\newblock


\bibitem[Rombach et~al\mbox{.}(2022)]%
        {DBLP:conf/cvpr/RombachBLEO22}
\bibfield{author}{\bibinfo{person}{Robin Rombach}, \bibinfo{person}{Andreas Blattmann}, \bibinfo{person}{Dominik Lorenz}, \bibinfo{person}{Patrick Esser}, {and} \bibinfo{person}{Bj{\"{o}}rn Ommer}.} \bibinfo{year}{2022}\natexlab{}.
\newblock \showarticletitle{High-Resolution Image Synthesis with Latent Diffusion Models}. In \bibinfo{booktitle}{\emph{{CVPR}}}. \bibinfo{publisher}{{IEEE}}, \bibinfo{pages}{10674--10685}.
\newblock


\bibitem[Ronneberger et~al\mbox{.}(2015)]%
        {DBLP:conf/miccai/RonnebergerFB15}
\bibfield{author}{\bibinfo{person}{Olaf Ronneberger}, \bibinfo{person}{Philipp Fischer}, {and} \bibinfo{person}{Thomas Brox}.} \bibinfo{year}{2015}\natexlab{}.
\newblock \showarticletitle{U-Net: Convolutional Networks for Biomedical Image Segmentation}. In \bibinfo{booktitle}{\emph{{MICCAI} {(3)}}} \emph{(\bibinfo{series}{Lecture Notes in Computer Science}, Vol.~\bibinfo{volume}{9351})}. \bibinfo{publisher}{Springer}, \bibinfo{pages}{234--241}.
\newblock


\bibitem[Sarkar et~al\mbox{.}(2023)]%
        {DBLP:conf/wacv/SarkarBVLBLM23}
\bibfield{author}{\bibinfo{person}{Rohan Sarkar}, \bibinfo{person}{Navaneeth Bodla}, \bibinfo{person}{Mariya~I. Vasileva}, \bibinfo{person}{Yen{-}Liang Lin}, \bibinfo{person}{Anurag Beniwal}, \bibinfo{person}{Alan Lu}, {and} \bibinfo{person}{Gerard Medioni}.} \bibinfo{year}{2023}\natexlab{}.
\newblock \showarticletitle{OutfitTransformer: Learning Outfit Representations for Fashion Recommendation}. In \bibinfo{booktitle}{\emph{{WACV}}}. \bibinfo{publisher}{{IEEE}}, \bibinfo{pages}{3590--3598}.
\newblock


\bibitem[Song et~al\mbox{.}(2021a)]%
        {DBLP:conf/iclr/SongME21}
\bibfield{author}{\bibinfo{person}{Jiaming Song}, \bibinfo{person}{Chenlin Meng}, {and} \bibinfo{person}{Stefano Ermon}.} \bibinfo{year}{2021}\natexlab{a}.
\newblock \showarticletitle{Denoising Diffusion Implicit Models}. In \bibinfo{booktitle}{\emph{{ICLR}}}. \bibinfo{publisher}{OpenReview.net}.
\newblock


\bibitem[Song et~al\mbox{.}(2017)]%
        {DBLP:conf/mm/SongFLLNM17}
\bibfield{author}{\bibinfo{person}{Xuemeng Song}, \bibinfo{person}{Fuli Feng}, \bibinfo{person}{Jinhuan Liu}, \bibinfo{person}{Zekun Li}, \bibinfo{person}{Liqiang Nie}, {and} \bibinfo{person}{Jun Ma}.} \bibinfo{year}{2017}\natexlab{}.
\newblock \showarticletitle{NeuroStylist: Neural Compatibility Modeling for Clothing Matching}. In \bibinfo{booktitle}{\emph{{ACM} Multimedia}}. \bibinfo{publisher}{{ACM}}, \bibinfo{pages}{753--761}.
\newblock


\bibitem[Song et~al\mbox{.}(2019)]%
        {DBLP:conf/mm/SongHLCXN19}
\bibfield{author}{\bibinfo{person}{Xuemeng Song}, \bibinfo{person}{Xianjing Han}, \bibinfo{person}{Yunkai Li}, \bibinfo{person}{Jingyuan Chen}, \bibinfo{person}{Xin{-}Shun Xu}, {and} \bibinfo{person}{Liqiang Nie}.} \bibinfo{year}{2019}\natexlab{}.
\newblock \showarticletitle{{GP-BPR:} Personalized Compatibility Modeling for Clothing Matching}. In \bibinfo{booktitle}{\emph{{ACM} Multimedia}}. \bibinfo{publisher}{{ACM}}, \bibinfo{pages}{320--328}.
\newblock


\bibitem[Song et~al\mbox{.}(2021b)]%
        {DBLP:conf/iclr/0011SKKEP21}
\bibfield{author}{\bibinfo{person}{Yang Song}, \bibinfo{person}{Jascha Sohl{-}Dickstein}, \bibinfo{person}{Diederik~P. Kingma}, \bibinfo{person}{Abhishek Kumar}, \bibinfo{person}{Stefano Ermon}, {and} \bibinfo{person}{Ben Poole}.} \bibinfo{year}{2021}\natexlab{b}.
\newblock \showarticletitle{Score-Based Generative Modeling through Stochastic Differential Equations}. In \bibinfo{booktitle}{\emph{{ICLR}}}. \bibinfo{publisher}{OpenReview.net}.
\newblock


\bibitem[Thanh{-}Tung and Tran(2020)]%
        {DBLP:conf/ijcnn/Thanh-Tung020}
\bibfield{author}{\bibinfo{person}{Hoang Thanh{-}Tung} {and} \bibinfo{person}{Truyen Tran}.} \bibinfo{year}{2020}\natexlab{}.
\newblock \showarticletitle{Catastrophic forgetting and mode collapse in GANs}. In \bibinfo{booktitle}{\emph{{IJCNN}}}.
\newblock


\bibitem[Tian et~al\mbox{.}(2023)]%
        {DBLP:conf/wacv/TianNB23}
\bibfield{author}{\bibinfo{person}{Yuxin Tian}, \bibinfo{person}{Shawn~D. Newsam}, {and} \bibinfo{person}{Kofi Boakye}.} \bibinfo{year}{2023}\natexlab{}.
\newblock \showarticletitle{Fashion Image Retrieval with Text Feedback by Additive Attention Compositional Learning}. In \bibinfo{booktitle}{\emph{{WACV}}}. \bibinfo{publisher}{{IEEE}}, \bibinfo{pages}{1011--1021}.
\newblock


\bibitem[van~den Oord et~al\mbox{.}(2018)]%
        {DBLP:journals/corr/abs-1807-03748}
\bibfield{author}{\bibinfo{person}{A{\"{a}}ron van~den Oord}, \bibinfo{person}{Yazhe Li}, {and} \bibinfo{person}{Oriol Vinyals}.} \bibinfo{year}{2018}\natexlab{}.
\newblock \showarticletitle{Representation Learning with Contrastive Predictive Coding}.
\newblock \bibinfo{journal}{\emph{CoRR}}  \bibinfo{volume}{abs/1807.03748} (\bibinfo{year}{2018}).
\newblock


\bibitem[Wang and Liu(2021)]%
        {DBLP:conf/cvpr/WangL21a}
\bibfield{author}{\bibinfo{person}{Feng Wang} {and} \bibinfo{person}{Huaping Liu}.} \bibinfo{year}{2021}\natexlab{}.
\newblock \showarticletitle{Understanding the Behaviour of Contrastive Loss}. In \bibinfo{booktitle}{\emph{{CVPR}}}. \bibinfo{publisher}{Computer Vision Foundation / {IEEE}}, \bibinfo{pages}{2495--2504}.
\newblock


\bibitem[Wang et~al\mbox{.}(2021)]%
        {DBLP:conf/icmcs/WangCWL21}
\bibfield{author}{\bibinfo{person}{Jianfeng Wang}, \bibinfo{person}{Xiaochun Cheng}, \bibinfo{person}{Ruomei Wang}, {and} \bibinfo{person}{Shaohui Liu}.} \bibinfo{year}{2021}\natexlab{}.
\newblock \showarticletitle{Learning Outfit Compatibility with Graph Attention Network and Visual-Semantic Embedding}. In \bibinfo{booktitle}{\emph{{ICME}}}. \bibinfo{publisher}{{IEEE}}, \bibinfo{pages}{1--6}.
\newblock


\bibitem[Wu et~al\mbox{.}(2019)]%
        {DBLP:journals/corr/abs-1908-08327}
\bibfield{author}{\bibinfo{person}{Jui{-}Chieh Wu}, \bibinfo{person}{Jos{\'{e}} Antonio~S{\'{a}}nchez Rodr{\'{\i}}guez}, {and} \bibinfo{person}{Humberto Jes{\'{u}}s~Corona Pamp{\'{\i}}n}.} \bibinfo{year}{2019}\natexlab{}.
\newblock \showarticletitle{Session-based Complementary Fashion Recommendations}.
\newblock \bibinfo{journal}{\emph{CoRR}}  \bibinfo{volume}{abs/1908.08327} (\bibinfo{year}{2019}).
\newblock


\bibitem[Zhan and Lin(2021)]%
        {DBLP:conf/icip/Zhan021}
\bibfield{author}{\bibinfo{person}{Huijing Zhan} {and} \bibinfo{person}{Jie Lin}.} \bibinfo{year}{2021}\natexlab{}.
\newblock \showarticletitle{{PAN:} Personalized Attention Network For Outfit Recommendation}. In \bibinfo{booktitle}{\emph{2021 {IEEE} International Conference on Image Processing, {ICIP} 2021}}. \bibinfo{publisher}{{IEEE}}, \bibinfo{pages}{2663--2667}.
\newblock


\bibitem[Zhang et~al\mbox{.}(2017)]%
        {DBLP:conf/iccv/ZhangXL17}
\bibfield{author}{\bibinfo{person}{Han Zhang}, \bibinfo{person}{Tao Xu}, {and} \bibinfo{person}{Hongsheng Li}.} \bibinfo{year}{2017}\natexlab{}.
\newblock \showarticletitle{StackGAN: Text to Photo-Realistic Image Synthesis with Stacked Generative Adversarial Networks}. In \bibinfo{booktitle}{\emph{{ICCV}}}. \bibinfo{publisher}{{IEEE} Computer Society}, \bibinfo{pages}{5908--5916}.
\newblock


\bibitem[Zhu et~al\mbox{.}(2017)]%
        {DBLP:conf/iccv/ZhuPIE17}
\bibfield{author}{\bibinfo{person}{Jun{-}Yan Zhu}, \bibinfo{person}{Taesung Park}, \bibinfo{person}{Phillip Isola}, {and} \bibinfo{person}{Alexei~A. Efros}.} \bibinfo{year}{2017}\natexlab{}.
\newblock \showarticletitle{Unpaired Image-to-Image Translation Using Cycle-Consistent Adversarial Networks}. In \bibinfo{booktitle}{\emph{{ICCV}}}. \bibinfo{publisher}{{IEEE} Computer Society}, \bibinfo{pages}{2242--2251}.
\newblock


\end{thebibliography}
